\documentclass[a4paper,11pt]{article}
\usepackage{graphicx}
\usepackage{xcolor}
\usepackage{amsmath,amsfonts,amssymb,amstext}
\usepackage[hidelinks]{hyperref}
\usepackage{hypcap}
\usepackage{multirow}
\usepackage{cancel}
\usepackage{empheq}
\usepackage{float}
\usepackage{cite}
\usepackage{subcaption}
\usepackage{authblk}
\usepackage{comment}
\usepackage{blindtext}
\usepackage{cases}
\usepackage{ulem}
\usepackage{epsfig}%
\usepackage[bottom]{footmisc}

\usepackage{epsfig}%
\usepackage[bottom]{footmisc}
\begin{document}
\date{}

\title{\centerline \bf  Dynamics of viable $f(R)$  dark energy models in the presence of Curvature-Matter interactions}
\bigskip

\author[1,2]{Anirban Chatterjee \thanks{Corresponding Author: anirban.chatterjee@gm.rkmvu.ac.in \& iitkanirbanc@gmail.com}}
\author[1]{Rahul Roy \thanks{rahulroyphysics@gmail.com}}
\author[1]{Sayantan Dey \thanks{sayantand84@gmail.com}}
\author[1]{Abhijit Bandyopadhyay\thanks{abhijit.phy@gm.rkmvu.ac.in}}
\normalsize
\affil[1]{Department of Physics, Ramakrishna Mission Vivekananda Educational and Research Institute, Belur Math, Howrah 711202, India}
\affil[2]{Department of Physics, Indian Institute of Technology Kanpur, Kanpur 208016, India}
\date{\today}
\maketitle

\begin{abstract}  
In this study, we analyze the dynamics of the interaction between dark matter and curvature-driven dark energy in viable $f(R)$ gravity models using the framework of dynamical system analysis. We incorporate this interaction by introducing a source term in their respective continuity equations, given by $Q = \frac{\kappa^2 \alpha}{3H}\tilde{\rho}_{\rm m}\rho_{\rm curv}$, and examine two $f(R)$ gravity models that comply with local gravity constraints and cosmological viability criteria.  Our findings reveal subtle modifications to fixed points and their stability criteria when compared to the conventional dynamical analysis of $f(R)$ gravity models without matter-curvature interactions as proposed and examined in prior literature. We determine the parameter limits associated with the stability criteria of critical points of the dynamical system  for both the models. Additionally, the introduction of interaction reveals variations in the dynamical aspects of cosmic evolution, contingent on the range of values for the relevant model parameters. These results are consistent with the observed features of cosmic evolution within specific limits of the $f(R)$ models parameters and the coupling parameter $\alpha$. We also examine the evolutionary dynamics of the universe in the interacting scenario through cosmological and cosmographic parameters. Our analysis demonstrates that the interacting scenario comprehensively accounts for all the observed phases of the universe's evolution and also results in a stable late-time cosmic acceleration. Additionally, we've studied how the coupling parameter affects evolutionary dynamics, particularly its impact on the matter-to-curvature energy density ratio. Our findings suggest that the chosen form of interaction can also address the cosmic coincidence problem.

\end{abstract}

\section{Introduction}
The measurement of redshift and luminosity distance for type Ia supernovae events  \cite{ref:Riess98, ref:Perlmutter}, played a pivotal role in confirming the transition of the universe from a decelerated phase to its current accelerated phase during its late-time evolution.
These findings are substantiated by the analysis of temperature anisotropies in the cosmic microwave background data from the WMAP mission \cite{WMAP:2003elm,Hinshaw:2008kr} and the detection of baryon acoustic oscillations \cite{SDSS:2005xqv}.
The general theory of relativity posits that the accelerated expansion is driven by a hypothetical energy density component characterized by negative pressure, commonly referred to as dark energy.
Investigating the origin and nature of dark energy responsible for the ongoing cosmic acceleration remains a prominent unsolved mystery in contemporary cosmology. 
The present-day cosmic acceleration cannot be accounted for by the standard equation of state, expressed as $\omega = p/\rho$, where $p$ and $\rho$
 represent the pressure and energy density of the conventional universe contents, such as radiation and matter.
In fact, we need a `yet-to-be-identified' component characterized by negative pressure, resulting in an equation of state $\omega < -1/3$, to provide an explanation for the accelerated expansion. \\

Initially, Einstein introduced a cosmological constant term $\Lambda g_{\mu\nu}$ into his field equations of general relativity to support his static universe theory. However, he later abandoned it in favor of Hubble's observations of an expanding universe. Nevertheless, in the latter part of the last century, as late-time cosmic acceleration was discovered, the cosmological constant term regained attention due to its potential to offer a straightforward solution to Einstein's field equations, resulting in an accelerated expansion. This gave rise to the $\Lambda$-CDM model, where `CDM' refers to cold dark matter. Unfortunately, this model grapples with the coincidence problem \cite{Zlatev:1998tr} and the fine-tuning problem \cite{Martin:2012bt}. These issues drive the exploration of alternative models to account for dark energy.\\

One category of models involves field-theoretic approaches to dark energy, where modifications are made to the energy-momentum tensor in Einstein's field equations due to the presence of a scalar field as an additional component in the universe, distinct from matter and radiation. These models encompass quintessence \cite{Peccei:1987mm,Ford:1987de,Peebles:2002gy,Nishioka:1992sg, Ferreira:1997au,Ferreira:1997hj,Caldwell:1997ii,Carroll:1998zi,Copeland:1997et} as well as $k-$essence models \cite{Fang:2014qga, ArmendarizPicon:1999rj,ArmendarizPicon:2000ah,ArmendarizPicon:2000dh,ArmendarizPicon:2005nz,Chiba:1999ka,ArkaniHamed:2003uy,Caldwell:1999ew,Bandyopadhyay:2017igc,Bandyopadhyay:2018zlz,Bandyopadhyay:2019ukl,Bandyopadhyay:2019vdd,Chatterjee:2022uyw}.
Another class of models focuses  on modifying the geometric aspects of Einstein's equations, particularly the Einstein-Hilbert action, as a means to address late-time cosmic acceleration. These models include $f(R)$ gravity models, scalar-tensor theories, Gauss-Bonnet gravity, and braneworld models of dark energy, as detailed in references \cite{fr1,fr2,fr3,fr4,fr5,fr6,fr7}. In many of these scenarios, the universe is assumed to exhibit homogeneity and isotropy on large scales, described by the Friedmann-Lema\^itre-Robertson-Walker (FLRW) metric with a time-dependent scale factor denoted as $a(t)$ and a curvature constant. Some approaches also consider an inhomogeneous universe, where a perturbed FLRW metric represents the background spacetime.\\

 Dynamical systems techniques are powerful tools for exploring cosmic evolution in generic cosmological models, as well as for investigating specific cosmological solutions. Dynamical systems analysis has recently been  employed to assess the stability of various scalar field dark energy models in comprehensive reviews \cite{Chatterjee:2021ijw,Chatterjee:2021hhj,Hussain:2022osn,Bhattacharya:2022wzu,Hussain:2023kwk}, as well as in the context of $f(R)$ gravity \cite{fr8,fr9,fr10, Bertolami:2009cd,Gunzig:2000ce,Odintsov:2017tbc,Lu:2019hra,Odintsov:2018uaw,Bahamonde:2017ize,Bahamonde:2019urw,Leon:2012mt,Xu:2012jf,Leon:2009dt,Leon:2009rc,Kofinas:2014aka,Basilakos:2019dof}. A systematic and comprehensive exploration of dynamical systems within $f(R)$ gravity theories has been carried out in a series of research  endeavors documented in  \cite{Samart:2021viu,Amendola:2006kh,Amendola:2006eh,Amendola:2007nt} and  references therein.
  These studies aimed to identify and classify models that faithfully portray the 
  correct cosmological evolution among a diverse set of $f(R)$ gravity models.
   Consequently, they contributed to the construction and development of `cosmologically viable' $f(R)$ gravity models, characterized by trajectories in the dynamical system phase space that align with observed cosmic phenomena.
 These viable $f(R)$ gravity models also conform to local gravity constraints, as extensively discussed with examples in references \cite{fr8,fr9,fr10}.  
What makes these models appealing is their ability to provide cosmic acceleration without requiring a cosmological constant or the introduction of dark energy as an additional field component in the universe. Instead, dark energy is 
`curvature-driven' - the dynamics described by the modified field 
equations incorporate the function $f(R)$ of Ricci  scalar 
($R$), which possesses the potential to drive 
cosmic acceleration, even when facing stringent constraints from 
both local gravity tests and various observational limitations.\\

In this study, we explore a scenario in which curvature-driven dark energy interacts with (dark) matter, all within the framework of viable $f(R)$ gravity models. We investigate whether the inclusion of these interactions in viable $f(R)$ models has a significant impact on achieving accurate cosmological acceleration when compared to $f(R)$ models that do not include matter-curvature interactions.
In $f(R)$-gravity theories which maintain  a minimal coupling to matter
\cite{fr1,fr2,fr3,fr4,fr5}, the dynamics originate from an action composed of two components: a modified `geometric' part obtained by replacing $R$ with $f(R)$ in the Einstein-Hilbert action, and a `matter' component arising from the conventional constituents (matter and radiation) of the universe. The involvement of the metric $g_{\mu\nu}$ in the matter component ensures minimal coupling with the geometry.
 The modified field equations can be restructured to assume the appearance of the standard field equations, expressed as $G_{\mu\nu} = 8\pi G T^{\rm (tot)}_{\mu\nu}$,
 where $G_{\mu\nu}$ is the Einstein tensor and $T^{\rm (tot)}_{\mu\nu} \equiv T^{\rm (M)}_{\mu\nu} + T^{\rm (curv)}_{\mu\nu}$. 
 The component $T^{\rm (curv)}_{\mu\nu}$ is  determined by $f(R)$ and its derivatives, and vanishes for $f(R) = R$ reducing to Einstein gravity. $T^{\rm (M)}_{\mu\nu}$ is a modification obtained by scaling the usual energy-momentum 
 tensor $\tilde{T}_{\mu\nu}$ from Einstein's equation by the factor $[df(R)/dR]^{-1}$.
$T^{\rm (curv)}_{\mu\nu}$ effectively serves as a representation of the stress-energy tensor, simulating a fluid-equivalent of the curvature function, while $T^{\mu\nu}$ is the adjusted stress-energy tensor for `matter and radiation' in the presence of gravity modifications arising from $f(R)$.  
In FLRW spacetime background, the conservation of $T^{\rm (tot)}_{\mu\nu}$ leads to a continuity equation that expresses the conservation in terms of the energy density and pressure for the comprehensive fluid, comprising radiation, matter, 
and curvature components. 
The continuity equation for each individual sector may allow for the inclusion of source terms, as long as the overall continuity equation for the comprehensive fluid remains valid without any source term.  
In this study, we considered radiation to be decoupled from matter and curvature
and proposed that interactions between the matter and curvature components 
introduce a source term $Q$ in their respective continuity equations, 
with opposite signs.
 The term $Q$ serves as a measure of 
rate of energy exchange between the matter and curvature sectors
owing to their interaction. In prior research \cite{Samart:2021viu}, the expression for $Q$ has been determined on
the multiplicative function of matter density and Hubble parameter. In this subsequent analysis, we take a further step by formulating the interaction term as $Q = \frac{\kappa^2 \alpha}{3H}\tilde{\rho}_{\rm m} \rho_{\rm curv}$
where, $\alpha$ signifies the strength of the coupling between
 matter and curvature with $\tilde{\rho}_{\rm m}$ and  $\rho_{\rm curv}$
 as their respective energy densities.\\
 
 The incorporation of interactions between matter and curvature thus introduces the
  additional parameter   $\alpha$  in into the analytical framework. 
  This coupling parameter, in conjunction with the parameters defining the function $f(R)$, becomes intricately entwined within the evolution equations,  playing a substantial role in shaping the evolutionary dynamics in the matter-curvature 
coupled scenario.  
We  employed dynamical analysis technique as a method to explore this interacting curvature-matter scenario
in the context of viable $f(R)$-gravity models.
This process entailed setting up the representative autonomous equations in terms of  appropriately
defined dynamical variables describing the cosmic evolution in this context. The essential 
components of our study involved identifying the   fixed points of the system
and examining their stability using the technique of 
linear stability analysis. These are crucial for achieving a comprehensive understanding of critical aspects in cosmic evolution, particularly those influenced by the interactions between curvature and matter, as considered in the present framework of our study.
In this investigation, we've opted for two specific $f(R)$ models,
namely, 
the generalised $\Lambda$-CDM model:
$f(R) = (R^b - \Lambda)^c,\quad  c \geqslant 1, bc \approx 1 $
and the power law model: $f(R) = R - \gamma R^n, \quad \gamma>0$, $0<n<1$. 
Both models have the potential to drive cosmic acceleration while maintaining their cosmological viability.
In general, the fixed points and characterisation of their stability are dependent 
on the parameters involved in the analysis, such as $\alpha$, ($b$, $c$) or $n$. Additionally, values of 
the density parameters for various components and the equation of state for the overall fluid at these fixed points 
provide  crucial insights into   the distinctive critical phases associated with each point.
We obtain the parameter limits within which a given fixed point may exhibit stable, unstable, or saddle-like 
behaviour. We 
also examine whether it can serve as an acceleration solution while staying within the 
boundaries of various cosmological 
constraints, as elaborated in subsequent sections.\\

Finally, we explore how curvature-driven dark energy influences various phases of cosmic evolution 
within the framework of matter-curvature interactions. 
To achieve this, we examined the evolution of  matter-to-curvature energy density ratio (denoted as $r_{mc}$
in the text), as well as two additional cosmographic parameters: deceleration $(q)$ and jerk $(j)$. The evolution of $r_{mc}$ provides quantitative information about the  dominance of curvature over matter at different stages of cosmic evolution. On the contrary, the deceleration parameter, along with the jerk, provides a detailed account of the kinematic aspects of evolution.  These investigations shed light on the potential scenarios that may arise within the context of matter-curvature interactions.  The observed evolution of the deceleration and jerk parameters suggests that all the mentioned models ultimately converge toward the $\Lambda$-CDM model in the distant future.\\

 The paper is organised as follows: In Sec.\ [\ref{Sec:2}], we establish a theoretical framework that 
 addresses the interaction between the curvature and matter sectors within the context of a flat FLRW 
 metric. Sec.\ [\ref{Sec:3}] is dedicated to formulating the autonomous equations of the dynamical system, 
 which incorporates the interaction between matter and curvature-driven dark energy within the framework of $f(R)$ gravity. Within this section, we also discussed the two specific $f(R)$ 
 models considered for this study. We present the fixed points of the autonomous system for each model and discuss their stability characteristics. Within Sec.\ [\ref{Sec:4}], we examine the influence 
 of curvature-driven dark energy on different phases of cosmic evolution. 
 We summarize our conclusions   in Sec.\ [\ref{Sec:5}].

\section{Theoretical framework of interacting curvature-matter scenario in $f(R)$ gravity}
\label{Sec:2}
Within the framework of $f(R)$ theory of gravity that  incorporates minimal curvature-matter coupling, the action can be expressed as follows \cite{fr1,fr2} 
\begin{eqnarray}
\mathcal{S} = \frac{1}{2\kappa^2}\int d^4x\sqrt{-g}f(R) + \int d^4x~\mathcal{L}_M(g_{\mu \nu},\phi _M)\,,
\label{eq:b1}
\end{eqnarray}
where  $\kappa^2=8\pi G$, $f(R)$ is an arbitrary function of the Ricci scalar $R$ and ${\cal L}_{M}$ commonly referred to as the `matter' Lagrangian, represents the Lagrangian density associated with both the radiation and matter (baryonic and CDM) components of the universe.  The symbol $g$ represents the determinant of the space-time metric tensor $g_{\mu\nu}$.
In metric formalism, the variation of this action with respect to the
field $g_{\mu \nu}$  gives the  
 corresponding modified  field equation as 
\begin{eqnarray}
F(R) R_{\mu\nu}-\frac{1}{2}g_{\mu \nu}f(R) + g_{\mu \nu}\square F(R) - \nabla_{\mu}\nabla_{\nu} F(R)  = \kappa^2 \tilde{T}_{\mu\nu}^{(\rm M)}\,.
\label{eq:b2}
\end{eqnarray}
Here, $F(R) \equiv df/dR$, and $\tilde{T}_{\mu \nu}^{(\rm M)}$ is the  stress-energy  tensor 
corresponding to radiation and matter components given by 
\begin{eqnarray}
\tilde{T}_{\mu \nu}^{(\rm M)} \equiv -{2 \over \sqrt{-g}}{\delta(\sqrt{-g}{\cal
L}_M)\over \delta(g^{\mu\nu})} \,.
\label{eq:b3}
\end{eqnarray}
Eq.\  (\ref{eq:b2}) may be reconfigured as \cite{fr1}
\begin{eqnarray}
 G_{\mu\nu} \equiv R_{\mu\nu}- \frac{1}{2}Rg_{\mu \nu} 
 = \kappa^2 \left(T_{\mu\nu}^{(\rm M)} + T_{\mu\nu}^{(\rm curv)}\right) 
 \equiv \kappa^2 T_{\mu\nu}^{(\rm tot)} \,,
\label{eq:b4}
\end{eqnarray}
where, $G_{\mu\nu}$ is the Einstein tensor and
\begin{eqnarray}
\kappa^2 T_{\mu\nu}^{(\rm curv)} & \equiv & 
\frac{1}{F} 
\left[\frac{1}{2}(f-RF)g_{\mu \nu} +  \left(\nabla_{\mu}\nabla_{\nu}- g_{\mu \nu}\square \right)F 
\right] 
\label{eq:b5} \\
\mbox{and} \quad T_{\mu\nu}^{(\rm M)}
& \equiv &
\frac{1}{F} \tilde{T}_{\mu\nu}^{(\rm M)}\label{eq:b6} 
\end{eqnarray}

In its reformulated form, the modified field equation  (\ref{eq:b4}) 
 describes a  FLRW universe
 featuring a fluid  with a comprehensive energy-momentum tensor denoted as
 $T_{\mu\nu}^{(\rm tot)} \equiv T_{\mu\nu}^{(\rm M)} + T_{\mu\nu}^{(\rm curv)}$. 
 The influence of the function $f(R)$ is readily apparent in both of these distinct components $T_{\mu\nu}^{(\rm curv)}$ and $T_{\mu\nu}^{(\rm M)}$, which collectively constitute $T_{\mu\nu}^{(\rm tot)}$.
 The component $T_{\mu\nu}^{(\rm curv)}$ is exclusively determined by 
 $f(R)$ and its higher derivatives and it vanishes for $f(R)=R$.
 This component effectively acts as a portrayal of the stress-energy tensor, mimicking a fluid-equivalent of the curvature function $f(R)$.  
 The component $T_{\mu\nu}^{(\rm M)}$ is obtained by modifying 
 $\tilde{T}_{\mu\nu}^{(\rm M)}$ using a functional multiplier of $1/F$ 
 (which equals 1 when $f(R) = R)$. 
 It serves as the  revised stress-energy tensor
 for `matter and radiation' in   presence of gravity modifications resulting from $f(R)$.
In the context of a flat FLRW spacetime background, when we regard the matter and radiation content as a perfect fluid, the (unmodified) stress-energy tensor $\tilde{T}_{\mu\nu}^{(\rm M)}$ becomes diagonal with entries being 
completely determined by the combined energy densities 
$(\tilde{\rho}_{\rm m} + \tilde{\rho}_{\rm r})$
of  dark matter ($\tilde{\rho}_m$),
radiation ($\tilde{\rho}_{\rm r}$), and the radiation pressure ($\tilde{P}_{\rm r} = \tilde{\rho}_r/3$). The   matter component  
is considered to be   non-relativistic pressureless dust. 
Consequently, the modified stress energy tensor  $T^{(M)}_{\mu\nu}  = (1/F)\tilde{T}_{\mu\nu}^{(\rm M)}$ corresponds to a fluid with energy density 
$(\rho_{\rm m} + \rho_{\rm r})$ and pressure $P_{\rm r}$, 
where $(\rho_{\rm r},\rho_{\rm m}, P_{\rm r})  \equiv (\tilde{\rho}_{\rm r}/F, \tilde{\rho}_{\rm m}/F, \tilde{P}_{\rm r}/F)$.
Note that, $\rho_i, P_{\rm r}$ ($i=\rm{r},\rm{m}$)   are always positive
because of the inherent positivity of ($\tilde{\rho_i},\tilde{P_{\rm r}}$)
and positivity of $F$ as well - a necessary condition for any cosmologically viable
 $f(R)$ model (discussed later in Sec.\ [\ref{Sec:3.2}]).
However, under such considerations,
the `00' and `$ii$' components of the modified field equation
 (\ref{eq:b4}) respectively yield  the following modified 
versions of the Friedmann equations; 
\begin{eqnarray}
\left(\frac{\dot{a}}{a}\right)^2   &\equiv & H^2 
= \frac{\kappa^2 }{3}\Big{[}  \rho_{\rm r}  + \rho_{\rm m} +   \rho_{\rm curv}  \Big{]}\,,
\label{eq:b7}\\
\frac{\ddot{a}}{a} &=& \dot{H} +  H^2 = - \frac{\kappa^2}{6}
\Big{[}    \rho_{\rm r}  + \rho_{\rm m}  + \rho_{\rm curv}  +   3(P_{\rm r}  + P_{\rm curv})\Big{]} 
\label{eq:b8}\,,
\end{eqnarray}
where $a$ is the FLRW scale factor, $H \equiv \dot{a}/a$ 
 (the Hubble parameter) and
dot $(\cdot)$   denotes   derivative  with respect to  
cosmic time. The quantities  
  $ \rho_{\rm curv}$ and $P_{\rm curv}$
 are defined through the following equations
 \begin{eqnarray}
\kappa^2 \rho_{\rm curv} & \equiv &  
 \frac{1}{F}\left(\frac{RF-f}{2}-3H\dot{R}F'\right)
\label{eq:b9}\\
\kappa^2 P_{\rm curv}& \equiv &\frac{1}{F}\left(\dot{R}^2F'' + 2H\dot{R}F'+\ddot{R}F' +\frac{1}{2}(f-RF)\right)
\label{eq:b10}\,,
\end{eqnarray}
where $'$ denotes derivative with respect to $R$.
It follows from eq.\ (\ref{eq:b5}) that, in a flat FLRW spacetime, the stress-energy tensor $T_{\mu\nu}^{\rm {curv}}$ representing curvature-fluid can be equated with that of an ideal fluid with  energy density and pressure given by $\rho_{\rm curv}$ and $P_{\rm curv}$ as provided in eqs.\ (\ref{eq:b9}) and (\ref{eq:b10}). Also note that, we
may express eq.\ (\ref{eq:b7}) as
\begin{eqnarray}
\Omega_{\rm m} + \Omega_{\rm r} + \Omega_{\rm curv} &=& 1
\label{eq:b11}
\end{eqnarray}
where $\Omega_{\rm m}$, $\Omega_{r}$ and $\Omega_{\rm curv}$ are modified 
form of density
parameters defined by following equations:
\begin{eqnarray}
\Omega_{\rm m} \equiv \frac{\kappa^2\rho_{\rm m}}{3H^2} = \frac{\kappa^2\tilde{\rho}_{\rm m}}{3FH^2}\,,\quad
\Omega_{\rm r} \equiv \frac{\kappa^2 \rho_{\rm r}}{3H^2} = \frac{\kappa^2\tilde{\rho}_{\rm r}}{3FH^2}\,,\quad
 \Omega_{\rm curv} \equiv \frac{\kappa^2\rho_{\rm curv}}{3H^2}
\label{eq:b12}
\end{eqnarray}
It's also worth mentioning that the specific case of
$f(R) = R$ 
($F=1, F',F''=0$) leads to
$\rho_{\rm curv} = 0$,
$P_{\rm curv} = 0$,  
$(\rho_i, P_{\rm r}) =( \tilde{\rho}_i,  \tilde{P}_{\rm r} )$  
implying reduction of   modified Friedmann  eqs. \eqref{eq:b7} and \eqref{eq:b8}
 to conventional Friedmann equations. \\

Taking divergence of the both sides of eq. \eqref{eq:b4} and use of Bianchi's identity 
lead  to the   conservation equation of total stress-energy tensor
$T_{\mu\nu}^{\rm tot} = T^{(M)}_{\mu\nu} + T^{\rm curv}_{\mu\nu}$:
\begin{eqnarray}
\nabla^{\mu}  T_{\mu\nu}^{\rm tot} &=& \nabla^{\mu}  \left(T_{\mu\nu}^{(\rm M)} + T_{\mu\nu}^{(\rm curv)}\right) =0 
\label{eq:b13}
\end{eqnarray}
In  FLRW spacetime, eq.\ (\ref{eq:b13})
takes the form of a continuity equation of the all-inclusive
fluid with   energy density 
${\rho}_{\rm tot} \equiv   \rho_{\rm r}   + \rho_{\rm m} + \rho_{\rm curv} $
and pressure $P_{\rm tot} \equiv P_{\rm r} +  P_{\rm curv}$
as 
\begin{eqnarray}
\dot{\rho}_{\rm tot}+3H(\rho_{\rm tot} +P_{\rm tot}) = 0 
\quad \mbox{or } \quad
\dot{\rho}_{\rm tot}+3H{\rho}_{\rm tot}(1 + \omega_{\rm tot}) = 0\,,
\label{eq:b14}
\end{eqnarray}
where $\omega_{\rm tot} \equiv  P_{\rm tot}/\rho_{\rm tot}$ represents
the  `grand' equation of
state (EoS) parameter of the comprehensive (radiation+matter+curvature) fluid.
Though the combined stress-energy tensor
$T^{(\rm tot)}_{\mu\nu}$ remains conserved, 
it is not necessary for its individual components:
$T^{(\rm curv)}_{\mu\nu}$ 
and $T^{(M)}_{\mu\nu}$  
(having  radiation $T^{(\rm r)}_{\mu\nu}$  and  
matter $T^{(\rm m)}_{\mu\nu}$ as its further sub-components)
to be separately conserved. 
This feature provides scope for incorporating `interactions' between the matter and curvature sectors.\\

Considering the negligible contribution of the radiation 
 component during the 
late time stages of cosmic evolution, we assume absence of interactions between 
radiation 
and combined matter-curvature  sectors  while simultaneously preserving the possibilities 
of interactions between  matter and curvature sectors. 
With such considerations, we posit the conservation eq. (\ref{eq:b13}) as a system of equations: (I) $\nabla^{\mu} T_{\mu\nu}^{(\rm r)} = 0$  and
(II) $\nabla^{\mu} T_{\mu\nu}^{(\rm m)} = -\nabla^{\mu} T_{\mu\nu}^{(\rm curv)} \equiv - \mathcal{Q}{\nu} \neq 0$ which collectively, validate the overall conservation of the comprehensive tensor 
$T^{(\rm tot)}_{\mu\nu}$. In FLRW spacetime, 
(I) implies $\dot\rho_{\rm r} +4H\rho_{\rm r} = 0$ ($\rho_{\rm r} \sim a^{-4}$) and
equation set (II) corresponds to non-conservation equations of the form:
\begin{eqnarray}
\dot{\rho}_{\rm curv}+3H\left(\rho_{\rm curv}+P_{\rm curv}\right)&=& Q \label{eq:b15}\\
\dot\rho_{\rm m} +3H\rho_{\rm m} &=&-Q
\label{eq:b16}
\end{eqnarray}
with a source term $Q$,  which  typically signifies a 
time-varying function   characterizing the instantaneous rate of energy exchange between the curvature and  matter sectors, 
reflecting their interactions. 
 Interaction between dark matter and curvature-driven dark energy 
can either be introduced  at the Lagrangian level, guided by field theoretic considerations,
or at a phenomenological level by modelling the  non-zero source term $Q$
(in eqs.\ (\ref{eq:b15}),(\ref{eq:b16})) in temrs of suitable cosmological
quantities. We adopt the latter approach, chosing a specific form of $Q$
 given by  $Q =  \frac{\kappa^2 \alpha}{3H}\tilde{\rho}_{\rm m}\rho_{\rm curv}$,
 where the coupling parameter $\alpha$ is dimensionless maintaining the
  dimensional consistency in both eqs. (\ref{eq:b15}) and (\ref{eq:b16}). 
  In this context, we may mention that, in  a previous study \cite{Samart:2021viu},
  the impact of interaction was explored using a functional form of the source term that exclusively depended on the energy density of the matter sector, Hubble parameter, and coupling constant in a multiplicative form. 
  Expanding upon this framework, our chosen form of the source term involves the multiplication of the energy densities of both sectors   capturing the instantaneous effects of both sectors on the source term with a dimensionless coupling constant. Our analysis reveals that such an interacting term also has the potential to address the coincidence problem, as discussed in the concluding section of the paper. \\

We imposed the constraint $\rho_{\rm curv} > 0$  
  in  attributing the sense of `energy density' (of the fluid-equivalent of the curvature) to $\rho_{\rm curv}$. This constraint
 restricts $f(R)$ models to   adhere to the condition
$RF - f>6H\dot{R}F'$
as evident from eq.\ (\ref{eq:b9}). 
It also
 constraints the  coupling parameter $\alpha$
as $\rho_{\rm curv}$ is interconnected with it through the source term $Q$
 as described by  eq.\ (\ref{eq:b15}). 
Under such considerations,
each of the modified
density parameters $\Omega_{\rm r}$, $\Omega_{\rm m}$ and $\Omega_{\rm curv}$
remain positive and collectively subject to the constraint given by 
Eq.\ (\ref{eq:b11}). 
This allows us to navigate and explore the parameter space relevant to the scenario of curvature-matter interaction by varying the parameter $\Omega_{\rm m}$ within the range $0 \leqslant \Omega_{\rm m} \leqslant 1$, eventually leading to obtaining constraints on the parameters of the model.\\

The  acceleration phase ($\ddot{a} > 0$) corresponds to 
$\omega_{\rm tot} = P_{\rm tot}/\rho_{\rm tot} < -1/3$  
as indicated by eq.\ (\ref{eq:b8}).  $F(R)$ and its derivatives 
are involved in  $\omega_{\rm tot}$  through $\rho_{\rm curv}$ and $P_{\rm curv}$
which, in turn are linked to the coupling parameter 
$\alpha$  via Equations (\ref{eq:b15}) and (\ref{eq:b16}).
The values of the parameter $\alpha$ and
those involved in the parametrisation of $F(R)$
that satisfy the condition $\omega_{\rm tot} < -1/3$ during the  late-time   phase of cosmic evolution corresponds a `Dark Energy' scenario driven by   matter-curvature interactions. Non-phantom dark energy scenarios adhere to the constraint $-1 < \omega_{\rm tot} < -1/3$.

\section{Dynamical analysis of  $f(R)$ models with curvature-matter interactions}
\label{Sec:3}
\subsection{Autonomous equations of dynamical system}
\label{Sec:3.1}
We employ dynamical analysis  as a tool to 
examine the curvature-matter interaction scenario 
 within the framework of cosmologically viable $f(R)$ gravity models.
 In this subsection we furnish the set of autonomous equations
 in terms of 
 suitably constructed basic   dynamical variables
which serve to  characterize the cosmological evolution within the context under consideration. 
This sets the ground for the application of techniques of 
dynamical analysis   in our current context. \\

The basic dynamical variables are  \cite{fr10}
\begin{eqnarray}
X_{1}= - \frac{\dot F}{HF}, \quad  X_{2}= - \frac{f}{6FH^2}, \quad X_{3}=\frac{R}{6H^2},  \quad X_{4} =\frac{\kappa^2  \rho_{\rm r} }{3H^2} = \frac{\kappa^2  \tilde{\rho}_{r}  }{3FH^2}=\Omega_{r}
\label{eq:d0}
\end{eqnarray}
By introducing the dimensionless parameter denoted as $N = \ln a$ 
to address temporal variations and applying the given definitions of dynamical variables ($X_i$'s), 
we can express eqs.\ (\ref{eq:b9}), (\ref{eq:b15}), and (\ref{eq:b16}) 
 which capture the evolutionary dynamics within the curvature-matter interaction context, 
 as the following  set of 4 autonomous equations, forming a 4-D dynamical system:
\begin{eqnarray}
\frac{dX_{1}}{dN}&=& -1 - X_{1}X_{3}- X_{3} - 3X_{2} + X_{4} + X_{1}^2 \nonumber\\
&&\qquad +\ \alpha \left(1- X_{1} - X_{2} - X_{3} - X_{4}\right)
 \left(X_{1} + X_{2} + X_{3}\right) \label{eq:d1} \\
\frac{dX_{2}}{dN} &=& \frac{X_{1}X_{3}}{m}-X_{2}\left(2X_{3}-4-X_{1}\right)  \label{eq:d2}\\
\frac{dX_{3}}{dN} &=& -\frac{X_{1}X_{3}}{m} - 2X_{3}\left( X_{3}-2\right) \label{eq:d3}\\
\frac{dX_{4}}{dN} &=& - 2X_{3}X_{4} + X_{1}X_{4} \,, \label{eq:d4}
\end{eqnarray}
where
\begin{eqnarray}
m &\equiv & \frac{d \ln F}{d \ln R}= \frac{R F' }{F}\,,
\label{eq:d5}
\end{eqnarray}
and in this context another parameter $r$ is introduced as
\begin{eqnarray}
 r & \equiv & -\frac{d \ln f}{d \ln R}= -\frac{R F}{f} = \frac{X_3}{X_2}\,.
\label{eq:d6}
\end{eqnarray}
These dimensionless parameters  $m$ and $r$  both depend on $R$
through $f(R)$, allowing us 
 to write $m$ as a function of $r$ i.e. $m = m(r)$. Each specific functional relation  $m = m(r)$ corresponds to a distinct class of $f(R)$ models.
When the coupling parameter $\alpha$ is set to zero, the above set of equations   reduces to the autonomous equations corresponding to the scenario that does not take into account interactions between the curvature and matter sectors. 
Also note that, if $f(R)=R$, then the parameter $m$ (as well as $X_1$) goes to zero, rendering the construction of the dynamical system infeasible. \\

By virtue of eqs.\ (\ref{eq:b11}) and  (\ref{eq:b12})
the dynamical variables in eq.\ (\ref{eq:d0}) are
constrained by the equation
\begin{eqnarray}
\Omega_{\rm m}   = 1 - X_{1} - X_{2} - X_{3} - X_{4} 
\label{eq:d7}
\end{eqnarray}
%
where $\Omega_{\rm m}$
is subject to the inequality $0 \leqslant \Omega_{\rm m} \leqslant 1$ and we
can  additionally
write
\begin{eqnarray}
\Omega_{\rm curv}  =  1 - \Omega_{\rm m} -  \Omega_{r} =  X_{1} + X_{2} + X_{3}\,.
\label{eq:d8}
\end{eqnarray}
Also,  the  `grand'  EoS parameter $\omega_{\rm tot}$  
can be expressed as 
\begin{eqnarray}
\omega_{\rm tot} &=& -1 - \frac{2\dot{H}}{3H^2} = -\frac{1}{3}(2X_{3}-1)
\label{eq:d9}
\end{eqnarray}
 EoS parameter associated with the  solo curvature sector \cite{Basilakos:2019dof} can be written as, 
\begin{eqnarray}
\omega_{\rm curv} &=& \frac{P_{\rm curv}}{\rho_{\rm curv}} = -\frac{(2X_{3}+X_{4}-1) }{3(X_{1}+X_{2}+X_{3})}
\label{eq:d9a}
\end{eqnarray}

The ratio of matter  to curvature energy density, denoted as $r_{mc} \equiv \Omega_{\rm m}/\Omega_{\rm curv}$,
 along with two additional cosmographic parameters -   the deceleration $(q \equiv - a\ddot{a}/\dot{a}^2)$ 
 and the jerk $(j \equiv \dddot{a}/aH^3)$ - serves  as valuable indicators for evaluating various facets
  of cosmic dynamics.  
During late cosmic epochs, the ratio $r_{mc}$, approximately reflects the ratio 
of `dark matter' to `dark energy' densities, as observations suggest 
 baryonic contributions to be significantly smaller compared to dark matter contributions in the present-day universe. 
The parameter $r_{mc}$ thus holds the potential to address the issue
 of the apparent coincidence between 
dark matter and dark energy
densities in present-day universe (the cosmic coincidence problem). 
The deceleration ($q$) and jerk ($j$) parameters
directly describe the `kinematic' characteristics of cosmic evolution. 
A negative $q$  signifies an accelerating universe, which
 is a hallmark of dark energy dominance. The constant $q=-1$ value signifies presence of cosmological constant.  
A switchover from a positive to a negative   value of $q$ marks the transition point from a decelerating to an accelerating phase cosmic expansion. Furthermore, varying $q$ values correspond to dynamical dark energy models, indicating the evolving nature of the universe's acceleration.
The jerk parameter, $j$ is a measure of   rate of change of acceleration (i.e. of $q$) 
and thus captures even finer temporal details of cosmic acceleration.
Both $q$ and $j$   encapsulate kinematic aspects of cosmic evolution which leave  distinct imprints on the large-scale structure of the universe. 
Consequently, these parameters have a crucial role to play in our endeavours to understand the impact of dark energy across the different stages of cosmic evolution. Furthermore, these parameters are instrumental in facilitating comparisons and distinctions among different `dark energy models' that utilize various mechanisms to trigger cosmic acceleration. \\

In the context of the matter-curvature interaction scenario considered in this study, 
the expressions for the three parameters 
(\(r_{mc}, q, j\))  can be derived in 
 terms of the dynamical variables ($X_i$'s) as
\begin{eqnarray}
 r_{mc} &\equiv & \frac{\Omega_{\rm m}}{\Omega_{\rm curv}} 
 = \frac{1 - X_1 - X_2 - X_3 - X_4}{X_1 + X_2 + X_3} \label{eq:d10}\\
q &\equiv & - \frac{a\ddot{a}}{\dot{a}^2} = -1 - \frac{\dot{H}}{H^2} = 1 - X_3  \label{eq:d11} \\
j &\equiv & \frac{\dddot{a}}{aH^3} = -\frac{X_1 X_3}{m}+2 (1 - X_3)^2+(1 - X_3)-2 X_3 (X_3-2)  \label{eq:d12}
\end{eqnarray}
In Sec.\ \ref{Sec:3.2} we have computed  the temporal evolution of these three parameters
for a  FLRW universe.\\

The fixed points in the 4D dynamical system described above correspond to solutions  
for $X_i$'s of $dX_i/dN = 0, (i=1,\cdots,4)$).
These stationary solutions and the assessment of their stability are crucial
for a comprehensive grasp of critical aspects in cosmic evolution driven by curvature-matter interactions, as explored within the scope of this study.
To investigate the system's behaviour  around these critical points, 
we employ the linear stability analysis which involves performing a first-order Taylor expansion of the autonomous equations derived above, which have a generic expression of the form: $d\vec{X}/dN = \vec{f}(\vec{X})$, where $\vec{X} \equiv {X_1, \cdots, X_4}$ and $\vec{f} \equiv {f_1, \cdots, f_4}$ with $f_i$'s given
by right hand side of the set of autonomous equations 
derived in eqs.\ (\ref{eq:d1}, \ref{eq:d2}, \ref{eq:d3}, \ref{eq:d4}).
The Jacobian of the linear transformation $\vec{X} \to \vec{f}$
is the $4\times 4$  matrix
\begin{eqnarray}
\mathcal{J} &=& \vert\vert {\partial f_i}/{\partial x_j}\vert\vert\,.
\label{eq:d13} 
\end{eqnarray}
%
By examining the eigenvalues of  $\mathcal{J}$
computed at the critical points
 we may deduce the nature of the stability of the  fixed points
 and classify them accordingly.  
 When all the eigenvalues have   negative (positive) real parts, 
 the fixed point is classified as  asymptotically stable (unstable).
 On the other hand, if any pair of eigenvalues occur  with a relative
 opposite sign in their real parts, the corresponding fixed point is a
saddle point. 
However, if any of the eigenvalues approaches zero at the fixed point,
the linear stability theory proves inadequate, 
necessitating the application of 
center manifold theory for a more insightful exploration
 of the characteristics of these critical points.   However, in our analysis detailed in later sections, 
we did not find any non-hyperbolic fixed points within the 
scope of the various $f(R)$ models examined in this study.
This suffices for adhering to linear stability
analysis for the models considered here without necessitating an extended 
exploration using methods beyond linear stability analysis  \cite{Odintsov:2017tbc}.

\subsection{On the choice of $f(R)$-models}
\label{Sec:3.2}
The main objective of this study is to  investigate and  analyze the dynamics of
$f(R)$-driven dark energy models 
taking into account interactions between the curvature and matter sectors. 
To carry out this investigation, 
we have selected specific $f(R)$ models that can induce cosmic acceleration while maintaining their cosmological viability.
In the context of the metric formalism employed in this study, any $f(R)$ function that is cosmologically viable must satisfy a set of stringent conditions which have been comprehensively
discussed in  \cite{Faraoni:2008mf, Tsujikawa:2010sc}.
Below, we enumerate these conditions and provide the underlying rationale for each of them.
\begin{itemize}
\item[(i)] $F>0$ for $R\geqslant R_0 >0 $ :
Condition to avoid anti-gravity.
\item[(ii)] $F'>0$ for $R\geqslant R_0 >0$ :
Condition to avoid Dolgov-Kawasaki instability  (Alignment with local gravity tests).
\item[(iii)] $f(R)$ is close to ($R-2\Lambda)$ for $R \gg R_0$ :  Consistency with local gravity tests and
presence of matter dominated era.
\item[(iv)]  $0 < \frac{RF'}{F} < 1$  at $r = -\frac{RF}{f} = -2$   
: Stability of de-Sitter solution at late cosmic times.
\end{itemize}
Here $R_0$ represents the value of the
Ricci scalar at present epoch.  
The relationship  $m=m(r)$ between the dimensionless parameters $m$ and $r$, as 
outlined in eqs. (\ref{eq:d5}) and (\ref{eq:d6}), is determined by the choice of $f(R)$.
Conversely, a specific form of $m(r)$ corresponds to a specific $f(R)$. 
We investigate the autonomous system furnished in Sec.\ \ref{Sec:3.1},
focusing on two separate  viable scenarios: The first scenario relates to a  
$f(R)$ resulting in a constant $m$, while the second
concerns a form of $f(R)$ where $m$ is a function of $r$,
represented as $m(r)$. We designate these scenarios as (A) and (B)
and discuss their specifications below.

\begin{itemize}
\item[(A)]   
We understand that a scenario with a constant $m$ can be achieved through the modified gravity model given by 
$f(R) = (R^b - \Lambda)^c$ with $(c \geqslant 1, bc \approx 1)$. This model has been previously examined \textit{without} taking into account matter-curvature interactions in \cite{fr10, Amendola:2007nt}. 
This was proposed with the
aim of extending the 
$\Lambda$-CDM model to address local gravity constraints 
justifying it as a viable $f(R)$ model. 
 We will refer to this scenario as    
`generalised $\Lambda$-CDM model' or sometimes simply as `model-A' from here on.  
 When constrained to $c \geqslant 1$ and $bc \approx 1$, this model converges to the 
 $\Lambda$-CDM model with $m=0$, ensuring a viable cosmological evolution. 
 The $m$ and $r$ parameters for the model are given by
$m= \frac{(bc - 1)R^b - b\Lambda + \Lambda}{R^b - \Lambda}$,  $r=-\frac{bcR^b}{R^b-\Lambda}$
and consequently, the parameter $m$ can be expressed as $m(r)=\left(\frac{1-c}{c}\right)r+b-1$. A dynamical analysis (see \cite{fr10, Amendola:2007nt} for details)
of this model 
indicates that at stable points   $r = -1 - m$,
which results in $m=-1+bc$, a constant value.

\item[(B)] This scenario involves consideration of $m$
as a specific function  of $r$ given by $m(r) = \frac{n(1+r)}{r}$
which can be realised for the power law form $f(R) = R - \gamma R^n$ with  ($\gamma>0$, $0<n<1$).
We refer to this scenario in subsequent texts as `power law model' and sometimes for convenience 
as `model-B'. 
As comprehensively discussed in
 \cite{fr10,Li:2007xn}, the power law model  correctly describes  cosmic evolution  within
 framework of non-interacting curvature-matter scenarios and
  satisfies the condition for cosmological viability as well for
 the above mentioned range of $\gamma$ and $n$.
 This form of $f(R)$ gives $m=\frac{\gamma(n-1)nR^n}{R - \gamma n R^n}$
 and $r=\frac{R - \gamma n R^n}{R-\gamma R^n}$, 
 the elimination of $\gamma$ from which
  leads to the form $m(r) = \frac{n(1+r)}{r}$.

\end{itemize}

In the context of our study, choice of the above two cases: (A) $m=-1+bc=$ constant
and (B) $m = m(r) = \frac{n(1+r)}{r}$
are intended to
 explore implications of the two above mentioned $f(R)$-models
  in \textit{presence} of matter-curvature interactions.

\subsection{Stability analysis of fixed points in generalised $\Lambda$-CDM and power law models}
\label{Sec:3.3} 
In the context of model-A and model-B, we identify the fixed points of the autonomous system described in Sec.\ [\ref{Sec:3.1}]
representing curvature-matter interactions  within the framework of  $f(R)$ gravity.
The fixed points 
are obtained 
by solving the equations $dX_k/dN = 0$  ($k=1,2,3,4$), where $dX_k/dN$'s are given by eqs.\ (\ref{eq:d1},\ref{eq:d2},\ref{eq:d3},\ref{eq:d4}) with $m = -1+bc$ for model-A
and $m = m(r) = \frac{n(1+r)}{r} = \frac{n(X_2+X_3)}{X_3}$
for model-B.  The fact that $m$ is constant for model-A and $r$-dependent for model-B with $r =\frac{X_3}{X_2}$ 
i.e. expressible completely in terms dynamical
 variables $X_2$ and $X_3$, enables us to formulate the system as a closed set of four autonomous equations involving four dynamical variables and 
 the constant parameters ($\alpha, b, c, n$)
without the necessity for additional dynamical variables. 
\\

 The coupling parameter $\alpha$ serves as a common parameter in the analytical framework across both models due to the consideration of matter-curvature coupling in our analysis.
Moreover, the  $f(R)$ model   includes
distinct model parameters: $(b,c)$ for model-A and $(n)$ for model-B. 
We can also evaluate the values of density parameters ($\Omega_{\rm r}$,  $\Omega_{\rm m}$, $\Omega_{\rm curv}$), grand EoS parameter $\omega_{\rm tot}$  and EoS parameter associated with curvature sector $\omega_{\rm curv}$ at the fixed points using eqs.\ (\ref{eq:d0}), (\ref{eq:d7}), (\ref{eq:d8}),  (\ref{eq:d9}), (\ref{eq:d9a}).
As per our considerations  
mentioned in Sec.\ [\ref{Sec:2}]  each of the modified density parameters 
 are positive 
and subject to the constraint given by eq.\ (\ref{eq:b11}).  
Adherence to these constraints determines which fixed points hold cosmological significance.  Therefore,
providing a description of fixed points and their stability characteristics, which are dependent on model parameters, requires appropriate referencing of the relevant constraints within the model parameter space. Moreover, the  density parameter values   and the  EoS parameter value  
at each fixed point provide crucial insights into characterizing 
the distinct critical phase associated with that point.
If $\omega_{\rm tot} < -1/3$ for any fixed point, it 
represents cosmic acceleration. Specifically, if the value falls within the range of $-1< \omega_{\rm tot} < -1/3$, the fixed point corresponds to a `non-phantom' acceleration and  when $\omega_{\rm tot} = -1$, it corresponds to the de-Sitter acceleration solution. 
The density parameter values at any given fixed point provide indications of the predominant component during the cosmic phase represented by that specific fixed point. \\

A total of 10 (real) fixed points are identified in model-A, while model-B comprises 7 fixed points. To distinguish them within the text, we label them as follows: 
\begin{eqnarray*}
 \mbox{For scenario (A)} &:& \{P_1,P_2,P_3,P_4,P_{5A},P_{6A},P_{7A},P_{8A},P_{9A},P_{10A}\}\\
\mbox{For scenario (B)} &:& \{P_1,P_2,P_3,P_4,P_{5B},P_{6B},P_{7B}\} 
\end{eqnarray*}
These fixed points, along with the corresponding values of $\Omega_{\rm m}$ and $\omega_{\rm tot}$ for both models, are displayed in Tabs.\ [\ref{tab:1}], [\ref{tab:2}], and [\ref{tab:3}].  
The arrangement of distributing the fixed points across three tables employs a classification scheme to facilitate a comparison of the characteristics attributed to the fixed points in both models.\\

\begin{table}[!h]
\centering
\begin{tabular}{|c|c|c|c|c|}
\hline
 Fixed  & \multirow{2}{*}{($X_1,X_2,X_3,X_4$)} & \multirow{2}{*}{$\Omega_{\rm m} $} & \multirow{2}{*}{$\omega_{\rm tot}$} & \multirow{2}{*}{$\omega_{\rm curv}$} \\
points & & & & \\
\hline\hline
$P_{1}$ & ($-4$ , 5 , 0 , 0) & 0 & $\frac13$ & $\frac13$ \\
\hline
$P_{2}$ & (0 , $-1$ , 2 , 0) & 0 & $-1$ & $-1$ \\
\hline
\multirow{2}{*}{$P_{3}$} & \multirow{2}{*}{$\left( -4 , \frac{4\alpha - 3}{\alpha} , 0 , 0 \right)$} & \multirow{2}{*}{$\frac{\alpha + 3}{\alpha}$} & \multirow{2}{*}{$\frac13$} & \multirow{2}{*}{-$\frac{\alpha}{9}$} \\
&&&&\\
\hline
\multirow{2}{*}{$P_4$} & \multirow{2}{*}{$\left(0 , \frac{-2\alpha - 3}{\alpha} , 2 , 0\right)$} & \multirow{2}{*}{$\frac{\alpha + 3}{\alpha}$} & \multirow{2}{*}{$-1$} & \multirow{2}{*}{$\frac{\alpha}{3}$} \\
&&&&\\
\hline
\end{tabular}
\caption{Common fixed points of model-A and model-B. The values of $\Omega_{\rm m}$, $\omega_{\rm tot}$ and $\omega_{\rm curv}$ at these fixed points are also presented.}
\label{tab:1}
\end{table}

\begin{table}[!h]
\centering
\begin{tabular}{|p{1cm}|c|c|c|c|}
\hline
Fixed   & \multirow{2}{*}{($X_1,X_2,X_3,X_4)$} &  \multirow{2}{*}{ $\Omega_{\rm m}$}  & \multirow{2}{*}{$\omega_{\rm tot}$} & \multirow{2}{*}{$\omega_{\rm curv}$} \\
points   &   & &  &      \\
\hline
\hline
\multirow{2}{*}{$P_{5A}$}  & \multirow{2}{*}{ $\left(  \frac{1}{\alpha-1}, 0 , 0 , 0\right)$} & \multirow{2}{*}{ $\frac{\alpha-2}{\alpha-1}$}& \multirow{2}{*}{$\frac13$} & \multirow{2}{*}{$\frac13 (\alpha - 1)$}\\
&&&&\\
\hline
$P_{6A}$  &	($ 1 $ , $ 0 $  , $0$ , $0$ ) & $0$&  $\frac13$ &  $\frac13$\\ 
\hline
$P_{7A}$  &	($ 0 $ , $ 0 $  , $0$ , $1$ ) & $0$ & $\frac13$ &  undefine \\
\hline
\end{tabular}  
 \caption{Fixed Points of model-A that are independent of $f(R)$
 model parameters $(b,c)$ but do not serve as fixed points of model-B.
The values of $ \Omega_{\rm m}$, $\omega_{\rm tot}$  and $\omega_{\rm curv}$
at these fixed points are also presented.}
\label{tab:2}
\end{table} 

The 4  points ($P_1, P_2, P_3, P_4$) presented in Tab.\ [\ref{tab:1}]
serve as common fixed points in both model-A and model-B and they are also independent 
of the parameters of their respective $f(R)$-model parameters. Notably, 
the first two points  $P_1$ and $P_2$  are even independent of the coupling parameter $\alpha$
implying that   they would also persist  as fixed points in the dynamical systems associated with non-interacting curvature-matter scenarios, regardless of any applied modifications to the gravity function $f(R)$.
Fixed points $P_2$ and $P_4$ correspond to $\omega_{\rm tot} = -1$, indicating an accelerating de-Sitter solution. The $\alpha$-dependence of the other two points ($P_3$ and $P_4$) arises directly from the consideration of matter-curvature interaction in the analysis, and they would not serve as (finite) fixed points in scenarios that do not account for such interactions. This can be understood by observing that the values of $X_2$ associated with these two points diverge as the matter-curvature coupling ($\alpha$) approaches zero. It is worth noting that for a critical value of $\alpha = -3$, the points $P_3$ and $P_4$ converge to the points $P_1$ and $P_2$ respectively. \\

The 3 fixed points ($P_{5A}, P_{6A}, P_{7A}$) in model-A, listed in Tab.\ [\ref{tab:2}], 
appear to be independent of the corresponding $f(R)$ model parameters $(b,c)$ but do 
not function as fixed points in model-B, which pertains to a different $f(R)$ model. 
The emergence of these points can be attributed to the constancy of the value of $m=-1+bc$ 
within the dynamical equations for model-A, in contrast to the situation in model-B where $m = 
\frac{n(X_2+X_3)}{X_3}$ itself enters as a function of the dynamical variables, as discussed 
earlier. It is worth noting that, the fixed point $P_{7A}$ corresponds to a state characterized by complete radiation dominance ($\Omega_{\rm r} = 1, \omega_{\rm tot} = 1/3$)
and none of the three points listed in Tab.\ [\ref{tab:2}] are representatives of
cosmic acceleration.\\

The remaining fixed points ($P_{8A}, P_{9A}, P_{10A}$) in model-A, as well as ($P_{5B}, P_{6B}, P_{7B}$) in model-B, depend on the parameter(s) of their respective $f(R)$ models. They are collectively presented in Tab.\ [\ref{tab:3}]. Since the expressions for $\omega_{\rm tot}$ at these fixed points depend on their respective model parameters, they would correspond to non-phantom accelerating solutions only if appropriate parameter values exist for which the condition $-1 < \omega_{\rm tot} < -1/3$ is satisfied.  \\

\begin{table}[h]
\centering
\begin{tabular}{|c| l|l|}
\hline
\scriptsize{Fixed}   & \hspace{3cm}\multirow{2}{*}{($X_1,X_2,X_3,X_4)$}   
& \multirow{2}{*}{\hspace{1cm} $ (\Omega_{\rm m}, \omega_{\rm tot}, \omega_{\rm curv})$}\\ 
\scriptsize{points} & &\\
\hline
\hline
\multicolumn{3}{|c|}{model-A} \\
\hline
\hline
\multirow{2}{*}{\scriptsize{$P_{8A}$}}& \multirow{2}{*}{ \scriptsize{$\left( \frac{4-2 b c}{-1+2 b c},\frac{5-4
b c}{1-3 b c+2 b^2 c^2},\frac{b c (-5+4 b c)}{1-3 b c+2 b^2 c^2},0\right)$} }& 
 \multirow{2}{*}{  \scriptsize{ $\left( 0\,, \frac{1+7bc - 6b^2c^2}{3-9bc + 6b^2c^2}\,, \frac{1+7bc - 6b^2c^2}{3-9bc + 6b^2c^2} \right)$}}   \\
&&  \\
  \hline
\multirow{2}{*}{\scriptsize{$P_{9A}$}}& \multirow{2}{*}{\scriptsize{$\left(  \frac{2 (1-bc) (3 b c+2 \alpha (bc-1))}{\alpha(1-3bc)-2(1-\alpha) (bc)^2},  \frac{-3+4 b c+4\alpha (1-b c)}{\alpha(1-3bc)-2(1-\alpha) (bc)^2},  
    \frac{b c (3-4 b c+4 \alpha (bc-1))}{\alpha(1-3bc)-2(1-\alpha) (bc)^2},0\right)$}} & 
  \scriptsize{$ \Big{(}\frac{3+\alpha-(13+\alpha)bc  + (8+2\alpha)(bc)^2 }
{\alpha - 3\alpha bc - 2b^2c^2 + 2\alpha b^2c^2 },  $}\\
&& \hspace{5mm}\scriptsize{$-\frac{(-1+bc)(\alpha- 6bc + 6\alpha bc )}
{3(\alpha  - 3\alpha bc - 2(1+\alpha)(bc)^2)}$,}\\
&& \hspace{5mm}\scriptsize{$\frac{6 b c (-1+\alpha )+\alpha }{-9+30 b c}\Big{)}$}\\

\hline
\multirow{2}{*}{\scriptsize{$P_{10A}$}}& \multirow{2}{*}{ \scriptsize{$\left( 4-\frac{4}{b c}, \frac{2-2 b c}{b^2 c^2}, 2-\frac{2}{b c}, -\frac{2}{b^2 c^2}+\frac{8}{b c}-5 \right)$}} &  \multirow{2}{*}{\scriptsize{$\left(0 , \frac{4}{3 b c}-1, \frac{-1+b c}{-3+9 b c}\right)$}} \\ 
&& \\
\hline
\hline
\multicolumn{3}{|c|}{model-B} \\
\hline
\hline
 \multirow{2}{*}{\scriptsize{$P_{5B}$}} &   \multirow{2}{*}{\scriptsize{ $ \left(-\frac{2(-2+n)}{2n-1}, \frac{5-4n}{1-3n+2n^2}, \frac{n(4n-5)}{1-3n+2n^2}, 0\right)$}}   & 
 \multirow{2}{*}{\scriptsize{$\left(0\,, \frac{-6n^2+7n+1}{3(2n^2-3n+1) }, \frac{-6n^2+7n+1}{3(2n^2-3n+1) }\right)$}}\\
&&  \\
\hline
\multirow{2}{*}{\scriptsize{$P_{6B}$}} & \multirow{2}{*}{\scriptsize{ $\left(-\frac{2 (-1+n) (2 \alpha (-1+n)+3 n)}{\alpha-3 \alpha n-2 n^2+2 \alpha n^2},\frac{-3-4 \alpha (-1+n)+4 n}{\alpha-3 \alpha n-2 n^2+2 \alpha n^2} , \frac{(3+4
\alpha (-1+n)-4 n) n}{\alpha-3 \alpha n-2 n^2+2 \alpha n^2}, 0\right)$}} 
&\scriptsize{$\Big{(} \frac{(n-1)(\alpha(2n-1) + 10n-3) - 2n^2}{\alpha(n-1)(2n-1) -2n^2}, $}\\
&& \hspace{1cm}\scriptsize{$ - \frac{(n-1)(\alpha-6n+6n\alpha)}{3\alpha(n-1)(2n-1) -6n^2} , $ }\\
&& \hspace{1cm}\scriptsize{$ \frac{6 n (-1+\alpha )+\alpha }{-9+30 n} \Big{)} $ }\\
\hline
\multirow{2}{*}{\scriptsize{$P_{7B}$}} 
&\multirow{2}{*}{\scriptsize{$ \left(\frac{4 (n-1)}{n}, -\frac{2 (n-1)}{n^2} , \frac{2 (n-1)}{n}, \frac{-5 n^2+8 n-2}{n^2}\right)$}}
&\multirow{2}{*}{\scriptsize{$\left(0\,, \frac{4}{3 n}-1 , \frac{-1+n}{-3+9 n}\right)$}}\\
&&  \\
\hline
\end{tabular}
\caption{ Fixed points in model-A and model-B
that are dependent on parameters of corresponding $f(R)$ models. The 
values of $ \Omega_{\rm m} $, $\omega_{\rm tot}$ and $\omega_{\rm curv}$  
at these fixed points are also presented.}
\label{tab:3}
\end{table}

Regarding the fixed points identified in our analysis, it's worth noting that if we were to exclude matter-curvature interactions from our considerations by setting $\alpha = 0$, 
the resulting dynamical system would consist of 8 fixed points model-A and 5 fixed points for model-B, as detailed below.
\begin{eqnarray*}
 \mbox{for model-A} &:&  \{P_1,P_2,P_{5A}^\star,P_{6A},P_{7A},P_{8A},P_{9A}^\star,P_{10A}\}   \\
\mbox{for model-B} &:&   \{P_1,P_2,P_{5B},{P_{6B}}^\star,P_{7B} \}   
 \end{eqnarray*}
where the points $P_{5A}^\star = (-1,0,0,0)$,   
$P_{9A}^\star  = \left(  \frac{2 (1-bc) (3 b c)}{ -2 (bc)^2},  \frac{-3+4 b c  }{ -2  (bc)^2},  
\frac{b c (3-4 b c )}{ -2  (bc)^2},0\right)$
and ${P_{6B}^\star} = \left( \frac{2 (-1+n) ( 3 n)}{  2 n^2 },\frac{-3+4 n}{  -2 n^2 } , \frac{(3-4 n) n}{  -2 n^2 }, 0\right)$ follow from the respective points $P_{5A}, P_{9A}, P_{6B}$ (given in   Tabs.\ [\ref{tab:2}] and [\ref{tab:3}]) when we set $\alpha=0$.
However, point $P_{5A}^\star$ would entail a value of $\Omega_{\rm m}=2$, falling outside the  bound  of $0 < \Omega_{\rm m} < 1$.  \\

\begin{table}[!h]
\centering
\begin{tabular}{|l|p{14cm}|}
\hline
Fixed  &\multirow{2}{*}{\hspace{4.5cm} Stability \& characteristics }   \\
points   &   \\
\hline
\hline
$P_1$ &    Stable for  $\alpha >-3$ and $0<bc<1$, otherwise  saddle. It neither represents matter dominated epoch nor dark energy epoch. \\  
\hline
$P_2$ &   Spiral stable $\alpha>-3$ and  $1<bc<\frac{41}{25}$. This critical point reveals stable de-sitter acceleration. \\  
\hline 
$P_3$ &   Stable for $\alpha \leq-3$ and $0 < bc< 1$, otherwise saddle. This point exhibits similar behavior like $P_{1}$, no interesting cosmological sequences regardless the values of $\alpha$.\\
\hline
$P_4$ &   Stable for $-18 <\alpha < -3$ and $1<bc<1.65$. This point represents stable de -sitter acceleration within the coexistence of both matter and curvature components for any value of $\alpha$ except $\alpha=-3$. For $\alpha=-3$, this de-sitter attractor point is driven by the curvature only like $P_2$.\\
\hline
$P_{5A}$ &   Stable if $bc<1 $ and  $0.75<\alpha<1$,  unstable if $\alpha>2$. Otherwise, this point is saddle.\\
\hline 
$P_{6A}$ &  Unstable if $bc>4/3$ and $\alpha<-2$, otherwise saddle. No cosmic acceleration has been found here. This point shows pure kinetic dominance.\\   
\hline
$P_{7A}$ &  This critical point exhibits saddle type for any values of $\alpha$. This point cosmologically reveals as pure radiation dominance.\\
\hline 
$P_{8A}$ &   Stable acceleration (non-phantom) for parameter ranges as depicted  in fig.\ [\ref{fig:1}](a), which corresponds to a bound on coupling as $\alpha>-4$.\\   
\hline
$P_{9A}$ &    Stable acceleration (non-phantom) for  parameter ranges as depicted  in fig.\ [\ref{fig:1}](b), which corresponds to a bound on coupling as $\alpha \leq -4$. \\
\hline
$P_{10A}$ &  Displays a saddle-like behavior, lacking both stability and cosmic acceleration for any parameter values. In the limit as $bc \rightarrow 1$, it is purely dominated by radiation. Otherwise, all three density components $\Omega_{\rm r}$, $\Omega_{\rm m}$, and $\Omega_{\rm curv}$ are present.\\

\hline
\end{tabular}  
\caption{Stability and cosmological features of the fixed points of model-A }
\label{tab:4} 
\end{table}

\begin{figure}[H]
  \centering
  \begin{tabular}{cc}
    \includegraphics[width=0.49\textwidth, height=1.94in]{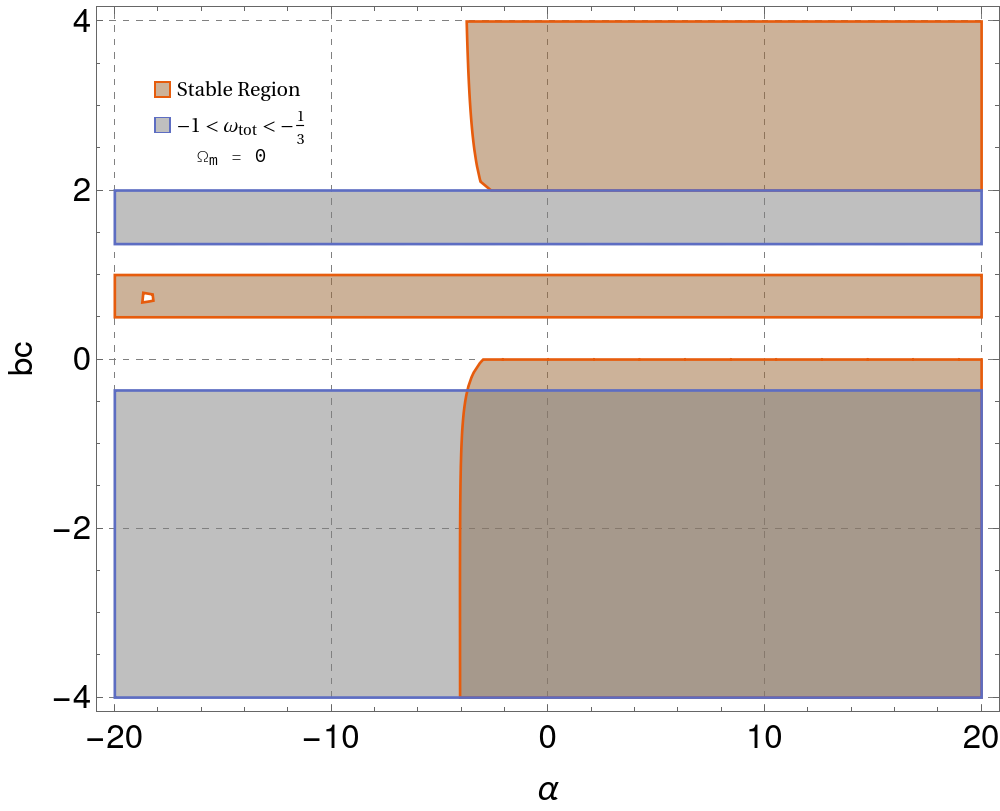} &
    \includegraphics[width=0.49\textwidth, height=1.94in]{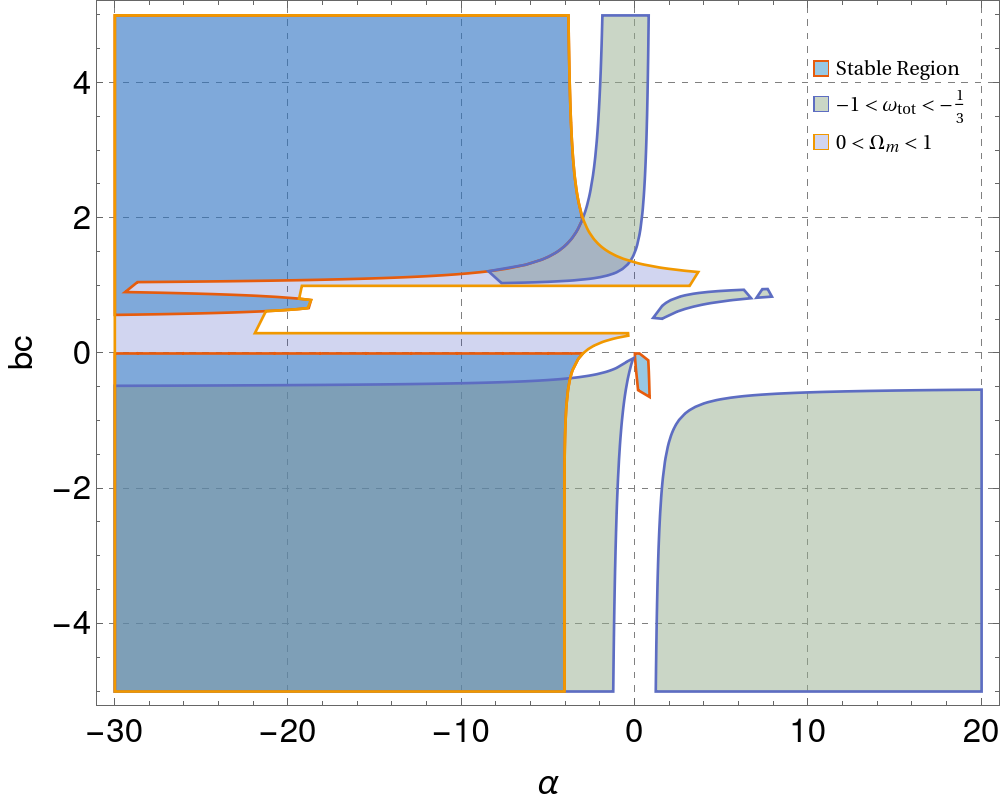} \\
    \textbf{(a) Point $\rm P_{8A}$} & \textbf{(b) Point $\rm P_{9A}$} \\[12pt]
  \end{tabular}
  
  
  \begin{tabular}{c}
    \includegraphics[width=0.49\textwidth, height=1.94in]{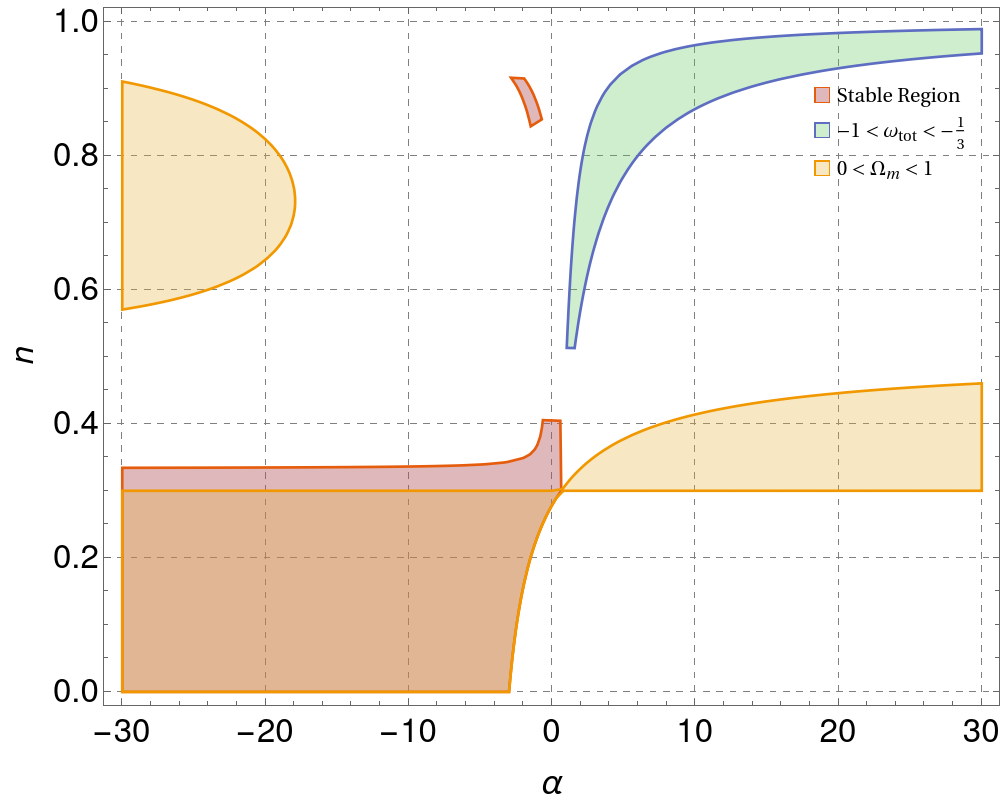} \\
    \textbf{(c) Point $\rm P_{6B}$} \\[6pt]
  \end{tabular}
  
  \caption{Allowed region of the parameter space for which the critical points \textbf{(a)} $P_{8A}$, \textbf{(b)} $P_{9A}$, and \textbf{(c)} $P_{6B}$ (i) exhibit stability, (ii) exhibit cosmic acceleration ($-1 < \omega_{\rm tot} < -1/3$), and (iii) satisfy the energy density condition $0<\Omega_{\rm m}<1$ are separately shown.}
  \label{fig:1}
\end{figure}

Note that, here we have examined the stability properties of fixed points of the relevant
autonomous system that occur with finite coordinate values. However, 
 when dealing with scenarios involving non-compact phase
spaces, it becomes crucial to explore the possibility of fixed points occurring at 
infinity and evaluate their stability criteria.
The methodology for investigating features in the asymptotic regime is extensively discussed in \cite{Leon:2012mt,Xu:2012jf}. One may observe that
all the dynamical variables of this system are bounded owing to the imposition of
of the various energy density constraints $(0 \leq \Omega_m, \Omega_r \leq 1)$, $-1 < \omega_{\rm tot} <1/3$  throughout
the evolutionary era from radiation to dark energy domination and viable $f(R)$ dark energy condition. This eliminates the necessity to explore fixed points at infinity.
We examine the stability  characteristics of the various fixed points using linear stability analysis, as outlined in Sec.\ [\ref{Sec:3.1}]. This entails an examination
of the eigenvalues of the Jacobian matrix $\mathcal{J}$ (Eq.\ (\ref{eq:d13})). The expression for these eigenvalues typically involves  the coupling parameter $\alpha$ and the parameters of the relevant $f(R)$ model. 
Therefore, to explore the stability properties of the fixed points, it becomes essential to examine the eigenvalues of $\mathcal{J}$ across the parameter space that is pertinent to the particular model being studied.   
Due to the
substantial and intricate analytical expressions of most eigenvalues for arbitrary
values of model parameters, we opt not to include their detailed expressions in the
article.
We perform a thorough scanning of the parameter space  allowing for a wide range of values for $\alpha$, $(b,c)$, while concurrently enforcing the constraint $0<n<1$, which is essential to uphold
 the cosmological viability of the power law model.
We obtain the (i) parameter space limitations under which a given fixed point  
demonstrates stable / unstable  or saddle-like behaviour. 
We also identify (ii) the parameter range within which the fixed point 
represents an acceleration solution of a specific kind - phantom/ non-phantom or de-Sitter. 
Simultaneously, we isolate (iii) the parameter  constraints that ensure that $\Omega_{\rm m}$ 
 remains within the range of $0 <\Omega_{\rm m}< 1$. \\
 
We perform a comprehensive scrutiny of the parameter constraints derived from the three 
schemes (i), (ii), and (iii) and offer a succinct overview of the stability characteristics 
and other cosmological aspects related to each of the fixed points. 
This summary, including description of relevant parameter ranges, is presented in  Tab.\ [\ref{tab:4}]
and [\ref{tab:5}] for model-A and model-B respectively.   
Moreover,  for visualization purposes, we specifically selected three fixed points, namely, $P_{8A}$, $P_{9A}$, and $P_{6B}$, to illustrate the constraints within the parameter space. 
In fig.\ [\ref{fig:1}], we present three distinct panels, each illustrating the permissible parameter ranges for individual points, wherein they separately (i) exhibit stability, (ii) demonstrate cosmic acceleration ($-1 < \omega_{\rm tot} < -1/3$), and (iii) satisfy the condition of $0 < \Omega_{\rm m} < 1$.
Note that,  the condition $0 < \Omega_{\rm m} < 1$ is not applicable for the case of point $P_{8A}$ as it corresponds to an $\Omega_{\rm m}$ value that is independent of the model parameters ($\Omega_{\rm m} = 0$ for $P_{8A}$,   see  Tab.\ [\ref{tab:3}]).
The areas of overlap between (i) and (ii) for point $P_{8A}$ and among (i), (ii), and (iii) for point  $P_{9A}$    define the specific parameter domains for these two points within which `stable cosmic acceleration is achieved through curvature-driven (non-phantom) dark energy'. No intersection of the  three domains corresponding to (i), (ii) and (iii) are found for  point $P_{6B}$.\\ 

\begin{table}[H]
\centering
\begin{tabular}{|l|p{14cm}|}
\hline
Fixed  &\multirow{2}{*}{\hspace{4.5cm} Stability \& characteristics }   \\
points   &  \\
   \hline
     \hline
$P_1$ &   For any value of the coupling parameter $\alpha$,
this point exhibits saddle-type behavior within the pre-defined range of $n$. This point neither represents matter dominated epoch nor dark energy epoch. \\
 \hline   
$P_2$ & This ponit is spiral stable if $\alpha>-3$ and $0<n<\frac{32}{25}$. Within the viable rage of $n$ this point signifies a stable accelerating de-Sitter solution, predominantly governed by curvature contribution. \\
 \hline
$P_3$ & This point exhibits simililar behavior like $P_{1}$. Saddle point behaviour with no interesting cosmological sequences regardless of the values of $\alpha$.\\
 \hline
$P_4$ &   This point is also stable for $ \alpha \leqslant -3$ within the viable range of $n$. de-sitter like acceleration can be realised in presence of both matter and curvature components for any value of $\alpha$ except $\alpha=-3$. For $\alpha=-3$, this de-sitter attractor point  has driven by the curvature only like $P_2$. \\
 \hline
$P_{5B}$ &   Stable for $\alpha>-3$,  but there is no common region in the parameter space of $n$ and $\alpha$  where stability and accelerating condition simultaneously satisfy within the viable range of $n$.\\ 
 \hline
$P_{6B}$ &   Stable for $\alpha<0.8$, but like $P_{5B}$ this points represents stable non-accelerarting fixed points  illustrated in fig.\ [\ref{fig:1}](c). But in $n\rightarrow 1$ limits this critical point embodies the conventional saddle matter dominated scenario. \\  
 \hline 
$P_{7B}$ & The system exhibits stable or saddle-like behavior, contingent on the value of $\alpha$ within the feasible range of $n$. In this viable range, $\omega_{\text{tot}}$ is greater than $\frac{1}{3}$. However, in the limit as $n$ approaches 1, it is solely dominated by radiation. Otherwise, there is no cosmologically interesting scenario has been observed here.\\
\\
\hline
\end{tabular}  
\caption{Stability and cosmological features of the fixed points of model-B}
\label{tab:5} 
\end{table}

It's important to note that, within the framework of $f(R)$-gravity scenarios featuring curvature-matter interactions, the fixed points $P_2$ and $P_4$ (which are common to both model-A and model-B), as well as fixed point $P_{8A}$ and $P_{9A}$ in model-A, are the only fixed points that represent  `stable acceleration solutions.' These solutions remain stable within the specified range of $\alpha$ values, as outlined in Tabs.\ [\ref{tab:4}] and [\ref{tab:5}]. Specifically, $P_2$ and $P_4$ correspond to stable de-Sitter attractors, while $P_{8A}$  and $P_{9A}$ correspond to stable acceleration driven by non-phantom dark energy. It's important to emphasize that there are no other points that exhibit both stability  and acceleration  while adhering to the constraint $0 < \Omega_{\rm m} < 1$  for any combination of parameter values.\\

\section{Evolutionary dynamics of curvature-matter coupled scenario in the context of cosmological and cosmographic variables}
\label{Sec:4}
In the context of $f(R)$ gravity models featuring matter-curvature interactions, we analyzed cosmic evolution across the entire timeline - from the radiation epoch to the matter epoch and finally to the late-time cosmological era. To comprehend the complete dynamics of the intertwined curvature-matter scenario, we introduced three cosmological parameters, namely the density parameters associated with curvature and matter ($\Omega_i$), the grand equation of state (EoS) parameter ($\omega_{\rm tot.}$), the ratio of matter density to curvature density ($r_{mc}$), and two crucial cosmographic parameters - deceleration ($q$) and jerk ($j$). These cosmographic parameters, previously introduced in eqs. (\ref{eq:d10}), (\ref{eq:d11}), and (\ref{eq:d12}), play a pivotal role in providing a quantitative characterization of the evolutionary progression.\\

\begin{figure}[h]
\centering
\begin{subfigure}[b]{0.49\textwidth}
\includegraphics[width=\textwidth]{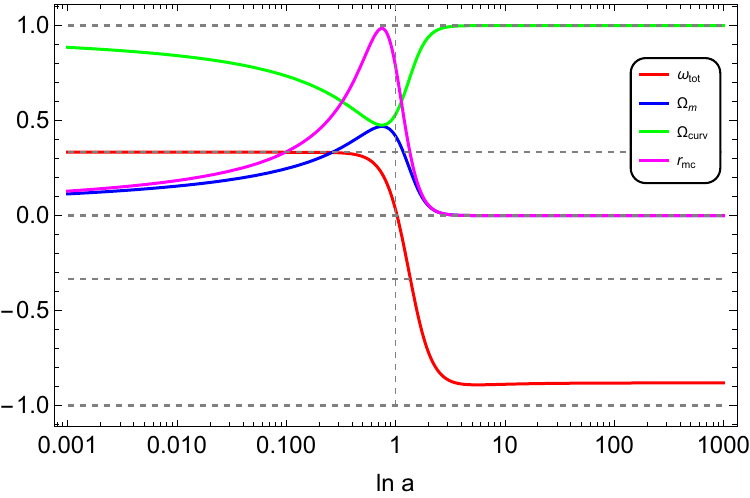}
\end{subfigure}
\begin{subfigure}[b]{0.475\textwidth}
\includegraphics[width=\textwidth]{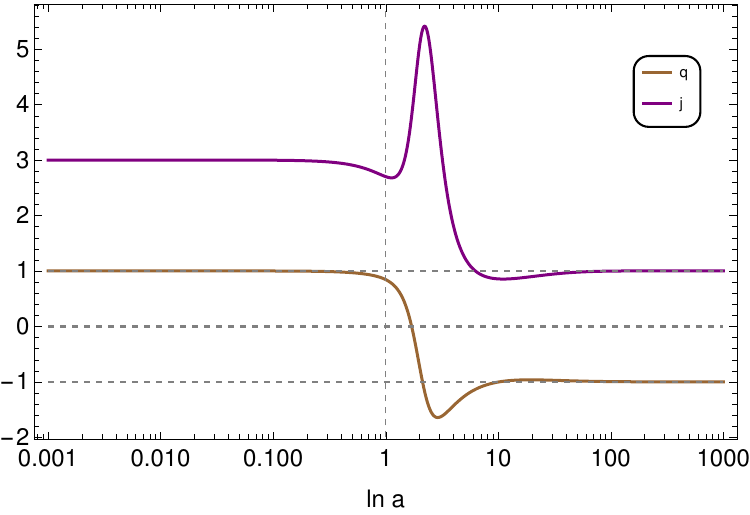}
\end{subfigure}
\caption{Evolution plots for generalised $\Lambda$-CDM model (scenario (A)).
Left Panel:  Plots of $r_{\rm mc}$, $\Omega_{\rm m}$, $\Omega_{\rm curv}$  
and $\omega_{\rm tot}$ vs $N (= \ln a)$ computed with   $bc = -3$ and  $\alpha=2$, 
Right panel: Plots of cosmographic parameters  $q$  and  $j$ 
computed with  $bc=1.5, \alpha=-2$.} 
\label{fig:2}
\end{figure}

\begin{figure}[H]
\centering
\begin{subfigure}[b]{0.49\textwidth}
 \includegraphics[width=\textwidth]{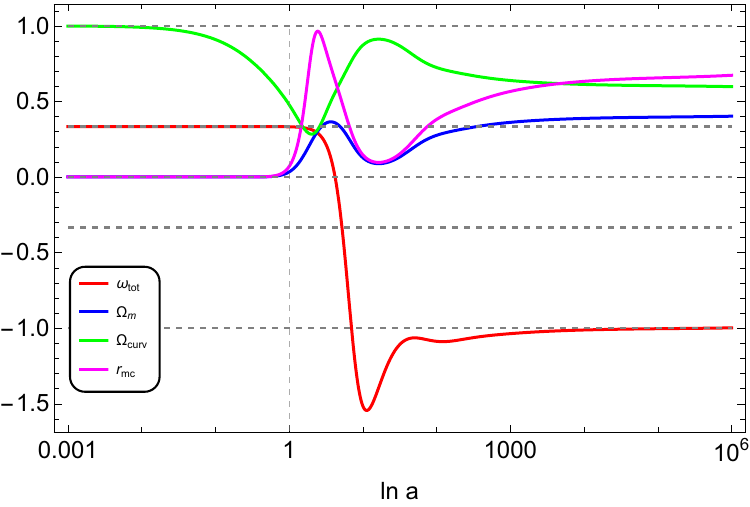}
\end{subfigure}
\begin{subfigure}[b]{0.49\textwidth}
\includegraphics[width=\textwidth]{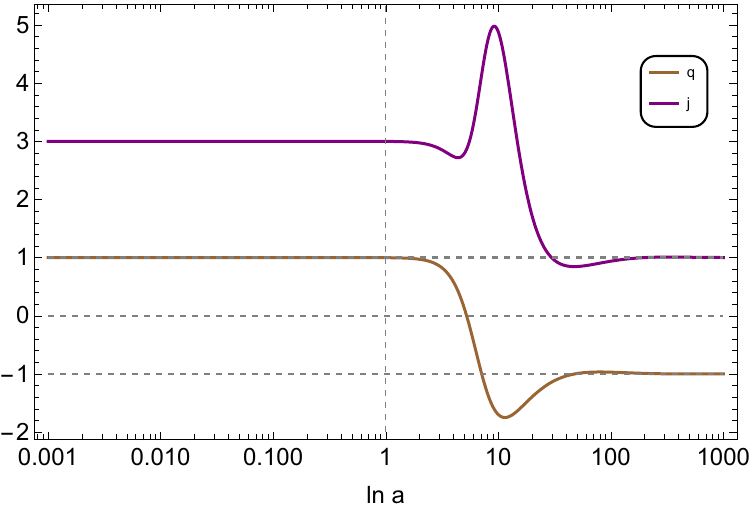}
\end{subfigure}
\caption{Evolution plots for power law model with $m=m(r)=\frac{n(1+r)}{r}$ (scenario (B)).
Left Panel:  Plots of $r_{\rm mc}$, $\Omega_{\rm m}$, $\Omega_{\rm curv}$ 
and $\omega_{\rm tot}$ vs $N (= \ln a)$ computed with  $n=0.9, \alpha=-5$, Right Panel: Plots of cosmographic parameters  
$q$  and  $j$ 
computed with  $n=0.9, \alpha=-2$.} 
\label{fig:3}
\end{figure}

 We numerically solve the system of autonomous eqs.\ (\ref{eq:d1}, \ref{eq:d2}, \ref{eq:d3}, \ref{eq:d4}) 
 for the dynamical variables  $X_i$ in the two cases - one with a constant $m$ 
corresponding to model-A and the other with
$m(r) = \frac{n(1+r)}{r}$, where $r = X_3/X_2$
corresponding to model-B.  Due to the non-linearity of the coupled autonomous equations, the solutions 
are highly sensitive to the initial  conditions necessary for solving this set of   equations.
However, to visually illustrate the potential evolutionary dynamics in  $f(R)$-gravity models with matter-curvature interactions, we establish specific valid boundary conditions that allow us to obtain
  solutions for $X_i$ based on carefully selected parameter values.
 These obtained solutions are then employed in equations (\ref{eq:d7}), (\ref{eq:d8}), and (\ref{eq:d9}) to compute the temporal profiles of density parameters ($\Omega_{\rm m}$, $\Omega_{\rm curv}$) and the equation of state parameter ($\omega_{\rm tot}$). Additionally, we determine the temporal evolution of the triad of parameters ($r_{mc}, q, j$) using eqs.\ (\ref{eq:d10}), (\ref{eq:d11}), and (\ref{eq:d12}). The depicted outcomes in fig.\ [\ref{fig:2}] model-A and fig.\ [\ref{fig:3}] for model-B demonstrate the evolution of $\Omega_{\rm m}$, $\Omega_{\rm curv}$, $\omega_{\rm tot}$, and $r_{mc}$ in the left panel, while showcasing the values of $q$ and $j$ in the right panel. This depiction covers a range of values for $N = \ln a$,  extending from 0.001 to 1000. The chosen parameter values ($\alpha$, $bc$ or $n$) for these representations are carefully selected to ensure that, they correspond to stable accelerating solutions and conform to the constraint $0 < \Omega_{\rm m} < 1$.  The specific parameter values are provided in the figure captions for their respective cases. \\

The graphs of $\Omega_{\rm curv}$, $\Omega_{\rm m}$, and $\omega_{\rm tot}$   
suggest the possibility of early cosmic phases where curvature energy density dominates over (dark) matter, yielding a positive $\omega_{\rm tot}$  (with the value of
 $\frac13$ for the cases presented in figs. [\ref{fig:2}] and [\ref{fig:3}]) and exhibiting a radiation-like EoS. As time progresses, curvature energy density diminishes while matter-energy density grows, and   their values tend to converge. At the matter-dominated epoch (where, $\omega_{\rm tot.}=0$), the values of the two densities become equal, triggering a shift in the nature of their temporal behaviour, resulting in the rise of curvature energy density and the decline of matter energy density. 
 It's worth noting that curvature energy density consistently exceeds matter energy density throughout the depicted evolutionary phases in the left panel of figs.\ [\ref{fig:2}] and [\ref{fig:3}]. 
 During moments when their values come together,  the equation of state parameter $\omega_{\rm tot}$, which had remained constant and positive since the initial epochs   experiences a rapid decline and swiftly 
 falls below $-\frac13$. This indicates the onset of late-time cosmic acceleration. For model (A), $\omega_{\rm tot}$ parameter can't cross the phantom barrier line, instead it saturates near the value of $-0.8$. As this phase unfolds, the values of $\Omega_{\rm curv}$ and $\Omega_{\rm m}$ once again separate, with curvature density regaining dominance over matter energy density.  
This dynamic facet of cosmic evolution is common to both scenarios (A) and (B), as can be inferred from figs.\ [\ref{fig:2}] and [\ref{fig:3}].\\

With a more in-depth comparison of figs.\ [\ref{fig:2}] and [\ref{fig:3}], one may also
recognize the distinctive impacts that the corresponding $f(R)$ modifications
 in model-A and model-B exert on the generic  features
of cosmic evolution mentioned above.
In model-A, as cosmic acceleration initiates, the rapid decline of $\omega_{\rm tot}$ does not cross the value of $-1$, but rather  gracefully saturates 
at a value  slightly above  $-1$, indicating a non-phantom nature of curvature-driven dark energy. 
In contrast, in model-B,  the sharp fall
of $\omega_{\rm tot}$   takes it below $-1$, but it subsequently rebounds and eventually settles at the value of $-1$, never exceeding it. 
 This feature implies that within the framework of the power-law model with matter-curvature interactions, the curvature-induced dark energy responsible for cosmic acceleration can exhibit characteristics of phantom dark-energy. Nevertheless, it's important to note that such a scenario does not exhibit stability
since $\omega_{\rm tot}$ eventually saturates -1 with the passage of time. 
Another notable distinction lies in the possibility
for late-time   cosmic acceleration to be purely curvature-dominated 
($\Omega_{\rm curv}=1$) in model-A. 
On the contrary, in model-B, cosmic acceleration can occur with both matter and curvature energy densities present in approximately equal proportions 
(on the order of $\sim$ 1), with an edge towards curvature dominance over matter.  \\

 Furthermore  in model-B, where $f(R)$ follows a power law form  indicated by $m(r) = n(1+r)/r$, as the universe progresses into its late-time acceleration phase, the temporal profiles illustrating $\Omega_{\rm curv}$ and $\Omega_{\rm m}$ display intricate and non-trivial patterns, rather than a simple, monotonic progression, until they eventually stabilize in the distant future. In contrast, these traits are absent in the constant $m$ 
  scenario (model-A). 
 This distinction can be attributed to the differences resulting 
 from the specific forms of $f(R)$ in each scenario. \\

\begin{figure}[t]
\centering

\begin{subfigure}[b]{0.48\textwidth}
\includegraphics[width=\textwidth]{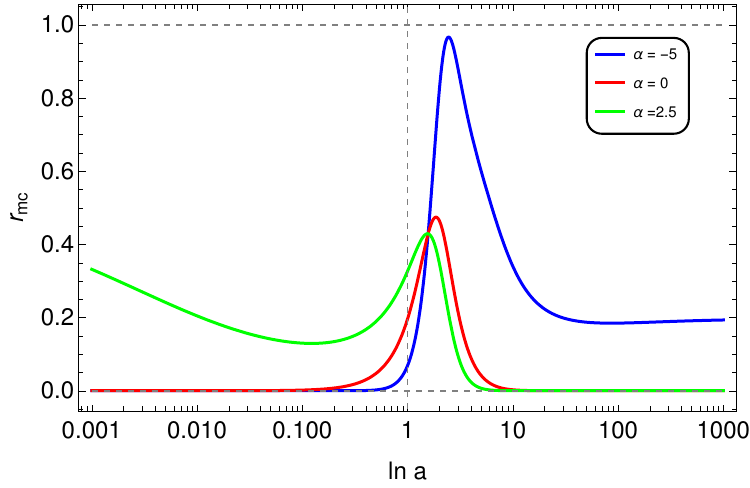}
\end{subfigure}
\begin{subfigure}[b]{0.48\textwidth}
\includegraphics[width=\textwidth]{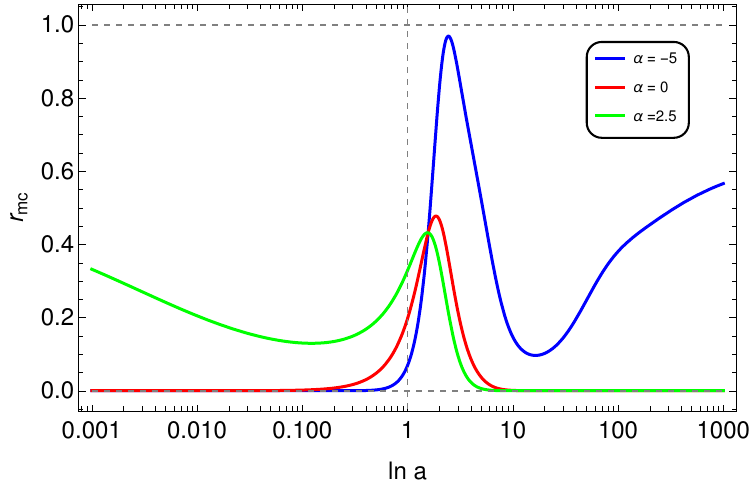}
\end{subfigure}
\caption{Evolution plot of the parameter $r_{\rm mc}$ for different values of 
coupling strength $\alpha$. Left Panel: For $f(R)$ model with generalised $\Lambda$-CDM 
model with parameter $bc=-3$. Right Panel: For power law form of $f(R)$ 
with   value of model parameter $n=0.9$.} 
\label{fig:4}
\end{figure} 

The ratio of matter to curvature density  $r_{\rm mc}  \equiv \Omega_{\rm m}/\Omega_{\rm curv}$,
 serves as a gauge of the extent to which one component dominates over the other.  
The value $r_{\rm mc} = 0$ signifies complete curvature dominance, while $r_{\rm mc} \approx 1$   suggests a coincidence of matter and curvature energy densities in the universe. 
Therefore, the evolution of the parameter $r_{\rm mc}$, as depicted in fig.\ [\ref{fig:2}] (model-A) 
and fig.\ [\ref{fig:3}] (model-B), provides quantitative insights into the dominance of curvature over matter at different stages of cosmic evolution
in the respective scenarios. 
The impact of the curvature-matter coupling strength on cosmic evolution becomes apparent when we compare the profiles of $r_{\rm mc}$ computed using different values of the interaction parameter $\alpha$. According to eqs. \eqref{eq:b15} and \eqref{eq:b16}, a negative value of $\alpha$ indicates energy transfer from the matter sector to the curvature sector, while a positive value of $\alpha$ implies the opposite, with energy flowing from curvature to matter. In fig. [\ref{fig:4}],  we display the evolution of the parameter $r_{\rm mc}$  for different $\alpha$-values in model-A (left panel)
  and model-B (right panel). These computations use specific values of $f(R)$-model parameters, 
as mentioned in the caption of fig.\ [\ref{fig:4}].  
Peaks in the $r_{\rm mc}$-profiles signify the moments when the universe attains its highest matter density proportion. 
The trend of the peak height approaching unity for $\alpha \sim -5$ or even more negative values implies the presence of epochs with almost perfect coincidence between curvature and matter energy densities  for significantly negative $\alpha$ values. 
For such negative $\alpha$ values, the late-time and distant-future accelerated phase of the universe experiences the presence of both matter and curvature energy densities in comparable proportions. Scenarios within the framework of the power law model (right panel of fig.\ [\ref{fig:4}]) may emerge for significantly negative $\alpha$ values where the matter energy density share may gradually rise as the universe progresses towards distant future epochs.
However,
 the $r_{\rm mc}$ values consistently remain below one, indicating that curvature density consistently exceeds matter energy density, even if by a slightly smaller margin during epochs when matter density contribution is at its highest.
At earlier epochs during the decelerated phase of the universe, the  negative $\alpha$ values result in very low $r_{\rm mc} \sim 0$ in both models, indicating curvature dominance over matter during this period. 
For the case of $\alpha = 0$  representing the standard $f(R)$
gravity framework without any additional
  matter-curvature interactions, 
the profile of $r_{\rm mc}$ exhibits a nearly symmetrical pattern
 centered  around the epoch of maximum matter-energy density contribution,
 with the matter energy density remaining exceedingly minimal ($r_{\rm mc} \sim 0$)
during earlier as well as during distant 
future epochs.
A positive $\alpha$, on the other hand,  results in a non-zero $r_{\rm mc}$ at early stages, which subsequently decreases over time, reaching a minimum before resuming its ascent to reach its highest point during the epoch when the universe maximizes its matter density share.\\

The right panels of figs.\ [\ref{fig:2}] and [\ref{fig:3}] depict the evolution of deceleration and jerk parameters ($q$ and $j$) for some benchmark values of the parameters corresponding to scenarios in model-A and model-B. These plots capture the kinematic aspects of cosmic evolution. During the early stages of evolution, they show positive values of both $q$ and $j$,  implying a decelerated phase with an increasing rate of deceleration. However, at a specific point, the deceleration parameter ($q$) rapidly declines, crosses the zero line, and becomes negative, signifying the onset of cosmic acceleration. It then sharply drops below the $-1$ threshold, rebounds, and ultimately stabilizes at $-1$ in the distant future.
This decline of $q$ and its eventual stabilization at $q=-1$ in the distant future are further characterized by a significant increase in the jerk parameter $j$, reaching its peak. This signifies a growing rate of acceleration, culminating in its maximum value. Subsequently, the jerk parameter starts to decrease, indicating a reduction in the rate of acceleration, and eventually settles at a positive value of unity in both model-A and model-B. All these intricate features of cosmic evolution result from the complex interplay among the interacting components of matter and curvature within the framework of modified gravity.

\section{Conclusion}
\label{Sec:5}
 
In this work, we explored the possibility for accommodating interactions 
between curvature and matter within the framework of viable $f(R)$ 
gravity models, all the while ensuring that these models retain their 
ability to achieve cosmic acceleration. 
At the action level, we consider minimal coupling between matter and curvature. 
The radiation component is considered to be decoupled from both matter and curvature, 
implying the conservation of its energy-momentum tensor. 
We incorporate interactions between the matter and curvature components by introducing a
 source term $Q$ in their respective continuity equations 
 with opposite signs.
 This ensures energy conservation within the combined matter-curvature sector, enabling the potential for a continuous exchange of energy between the two sectors at a rate determined by $Q$. 
A specific form of the interaction term $Q$ is chosen to illustrate a dynamic connection between the curvature and matter sectors by incorporating their individual energy densities in a multiplicative manner within $Q$, along with a dimensionless parameter ($\alpha$)
signifying the strength of coupling between the two sectors.
The sign of $\alpha$ dictates the direction of energy flow between the two sectors.\\

An essential component of this study is the establishment of
 a 4-D dynamical framework that  portrays the dynamics 
  of curvature-matter interacting scenario.
The introduction of the interaction term $Q$ leads to alterations 
in the autonomous equations that govern the dynamical system, 
distinguishing it from 
those resulting from $f(R)$ theories that do not consider matter-curvature interactions.
We identified the critical points of the dynamical system and analyzed their 
stability, separately in the context of two viable $f(R)$ models \textit{viz.} 
the generalized $\Lambda$-CDM model (model-A) and the Power-law model (model-B) . 
The interaction parameter $\alpha$ assumes a crucial role, significantly 
shaping the dynamics of the system.
Its impact  on the stability characteristics of various fixed points within the dynamical system, as well as  on the evolutionary dynamics of the universe during different phases, have been thoroughly investigated. \\

The incorporation of the curvature-matter coupling ($\alpha \neq 0$) 
within the framework of $f(R)$ gravity leads to an increase in 
the number of critical points, introducing 
new stable de-Sitter attractors.  
In both models, we notice the appearance of two stable de-Sitter attractors  $P_2$ and $P_4$, in contrast  
to  
 the scenario without interactions ($\alpha = 0$) where  $P_2$
 exists as the only de Sitter point.
 An in-depth analysis of the stability traits of these fixed points across an extensive range of $\alpha$-values reveals intriguing  patterns within the stability landscape of the matter-curvature interaction scenario. For $\alpha = -3$, the stable de-sitter attractors $P_2$ and $P_4$ represents the same fixed point in both models. In model-A, for $\alpha > -3$, $P_2$ takes the role
of stable de-Sitter attractor, whereas in the  range of $-18 < \alpha \leq -3$, 
it is $P_4$ that assumes this stabilizing role, provided  
the $f(R)$-model parameters $b$,$c$ satisfy $1 < bc < 1.65$.
On the other hand, in Model B, stable de-Sitter acceleration is achievable for $\alpha > -3$, with $P_2$ serving as the attractor and for $\alpha \leq -3$,  $P_4$  emerges as the stable de Sitter attractor, operating within the viable range of the $f(R)$ model parameter $0 < n < 1$. 
As evident from the above discussion, a significant contrast between the two models is that, in 
Model B, the   entire span of $\alpha$ permits stability of de-Sitter acceleration solution,
whereas in Model A, this range is notably more restricted.
Shifting our focus to identifying fixed points that signify stable acceleration  
of non-phantom nature ($-1<\omega_{\rm tot} < -1/3$), 
we observe that
model-A  presents two stable options, namely $P_{8A}$ and $P_{9A}$,
with their stability  demonstrated  for $\alpha > -4$ and $\alpha \leqslant -4$
respectively. 
Notably, these conditions are dependent on the parameters ($b$,$c$) of model-A, 
as illustrated in figs. [\ref{fig:1}](a) and [\ref{fig:1}](b). 
In contrast, within the viable range of $n$ in Model B, 
no stable fixed points with non-phantom acceleration have been found  for any value  
of $\alpha$.\\

The evolution of various cosmological and cosmographic parameters have been 
explored  
for a nuanced assessment of the  influences of curvature-driven dark energy
on
various phases of cosmic evolution within the framework of matter-curvature interactions.
Examining the evolution of modified energy densities and the EoS parameter, particularly the matter-to-curvature density ratio ($r_{\rm mc}$), along with two cosmographic parameters - deceleration ($q$) and jerk ($j$), offers valuable insights into potential evolutionary scenarios within scenarios involving the interplay of matter and curvature. 
In model-A, we observed that in both the early and late stages, 
curvature energy density exhibits significant dominance over matter energy density. 
During the transitional phase between these two phases, curvature density decreases 
while matter density increases over time, eventually reaching its peak. Subsequently, 
there is a resurgence in curvature's contribution and a decline in matter's contribution, 
ultimately reestablishing the pronounced dominance of curvature in the later stages. 
 During the transition from early to later phases, the EoS parameter swiftly drops below 
 $-1/3$, marking the onset of late-time cosmic acceleration and
 stabilizes around $\sim -0.8$ in distant future, never reaching below $-1$, affirming the non-phantom nature of curvature-driven dark energy.
   This stable acceleration phase in the distant future is achieved with prominent curvature dominance, indicated by $r_{\rm mc} \sim 0$.  
In model-B, while curvature density plays a dominant role in the early universe, 
matter density also becomes significant during the later phases, 
resulting in an order of magnitude around 1 (but always $<1$) 
for $r_{mc}$ in the later stages. 
The EoS parameter drops from its positive value to values below 
$-1/3$ as the universe enters its accelerated phase. 
The descent persists, surpassing the non-phantom threshold of $-1$ before rebounding and 
stabilizing at this value in the later stages signifing the emergence of an unstable phantom region in model-B, albeit for a brief period.  
The evolution of parameters ($q, j$) remains consistent across both models. 
The positive jerk ($j$) and negative deceleration ($q$) observed during the 
late-time phase in both models confirm the presence of 
stable late-time cosmic acceleration in interacting curvature-matter scenario.  
We examined and illustrated (fig. [\ref{fig:4}]) the progression of $r_{mc}$ at specific reference 
values of the coupling parameter $\alpha$. The parameter $r_{mc}$ is a measure of 
the extent to which matter dominates over the curvature component. In the absence of any interaction, the $r_{mc}$ profile remains symmetric around the epoch of maximum 
energy density contribution of the matter component, with $r_{mc}$ values tending 
to zero at both earlier and later time epochs in both models. 
When interactions are present ($\alpha \neq 0$), positive values of $\alpha$ result in earlier epochs showing a notable contribution from the matter component. Conversely, with negative $\alpha$ values, the later epochs display a substantial contribution from the (dark) matter component.
Our study is able to address the issue of cosmological coincidence in the late-time epoch with a negative $\alpha$. \\

  The ratio of matter density to curvature density, denoted as $r_{\rm mc}$ and defined as $\frac{\Omega_{\rm m}}{\Omega_{\rm curv}}$, serves as a measure of the dominance of one component over the other in the cosmic landscape.   $r_{\rm mc} = 0$ indicates complete curvature dominance, while $r_{\rm mc} \approx 1$ suggests a coincidental balance between matter and curvature energy densities in the universe. The evolution of the parameter $r_{\rm mc}$ is illustrated in fig.~[\ref{fig:2}] (model-A) and fig.~[\ref{fig:3}] (model-B), providing quantitative insights into the dominance of curvature over matter at different stages of cosmic evolution in the respective scenarios. The impact of the curvature-matter coupling strength becomes evident when comparing $r_{\rm mc}$ profiles computed with different values of the interaction parameter $\alpha$. Within the framework of equations \eqref{eq:b15} and \eqref{eq:b16}, a negative $\alpha$ implies energy transfer from matter to curvature, while a positive $\alpha$ signifies the opposite, with energy flowing from curvature to matter. fig.~[\ref{fig:4}] displays the evolution of $r_{\rm mc}$ for different $\alpha$ values in model-A (left panel) and model-B (right panel), using specific values of $f(R)$-model parameters as mentioned in the caption of fig.~[\ref{fig:4}]. Peaks in the $r_{\rm mc}$ profiles represent moments when the universe attains its highest matter density proportion. The trend of peak height approaching unity for $\alpha \sim -5$ or more negative values indicates epochs with nearly perfect coincidence between curvature and matter energy densities. For significantly negative $\alpha$ values, the late-time and distant-future accelerated phase of the universe experiences both matter and curvature energy densities in comparable proportions. However, $r_{\rm mc}$ values consistently remain below one, suggesting that curvature density consistently exceeds matter energy density, even during epochs with the highest matter density contribution. At earlier epochs during the decelerated phase of the universe, negative $\alpha$ values result in very low $r_{\rm mc}$, indicating curvature dominance over matter. In the case of $\alpha = 0$, representing the standard $f(R)$ gravity framework without additional matter-curvature interactions, the $r_{\rm mc}$ profile exhibits a nearly symmetrical pattern centered around the epoch of maximum matter-energy density contribution. A positive $\alpha$ results in a non-zero $r_{\rm mc}$ at early stages, decreasing over time and reaching a minimum before resuming its ascent to reach the highest point during the epoch when the universe maximizes its matter density share.\\

While numerous viable $f(R)$ models exist to address the enigma of dark energy 
in cosmology, the qualitative insights from our study can form a foundation for more intricate examinations. 
These findings serve as supplementary constraints when assessing the viability of
$f(R)$ models, alongside the broader range of cosmological 
constraints applied to $f(R)$ theories. 
Each $f(R)$ gravity model leaves its distinct imprint on
the evolution of cosmological perturbations. Specifically, the scrutiny of structure formation 
in the universe proves to be highly responsive to the nature of
interaction between dark energy and dark matter within the framework of $f(R)$ gravity. 
Notably, the interaction between curvature driven dark energy and dark matter 
in the early stages of the universe may potentially influence the epoch of matter-radiation equality. 
The manifestation of anisotropies resulting from this phenomenon can be 
quantitatively assessed in terms of the growth of structure formation within
the curvature-matter interacting framework of   $f(R)$ gravity. 
Interestingly, the interaction between
$f(R)$-dark energy and dark matter during the  
early phase of cosmological evolution may offer an explanation for the disparity in the Hubble tension parameter observed between  local measurements and those derived from cosmic microwave background observations. 
Furthermore, the study of matter perturbations and the local gravity approximations 
of interacting dark energy in $f(R)$ gravity holds promise for shedding 
light on the inconsistent predictions between non-interacting and interacting $f(R)$ 
dark energy models. 
Additionally, exploring alternative models of interacting dark energy within the 
framework of $f(R)$ gravity holds the potential to extract further physical 
implications and discern their cosmological consequences.\\

  In summary, this research deepens our understanding of the intricate dynamics in modified $f(R)$ gravity theories,
  particularly in the presence of interactions between curvature and matter. The emergence of stable de-Sitter 
  attractors, nuanced stability characteristics, and distinctive evolutionary patterns in this scenario are 
  instrumental in elucidating the various facets of the complex behavior of the system. 
The findings presented here not only enhance our understanding of the dynamics of these systems but also open up new avenues for further exploration in the field of cosmology, paving the way for gaining valuable insights into the fundamental dynamics of the universe.

\paragraph{Acknowledgement}\
We are thankful to the referees for their valuable comments.  A.C.   thanks    the
Department of Physics, IIT Kanpur for supporting this research through the Institute Post-Doctoral Fellowship (Ref.No.DF/PDF197/2020-IITK) at the initial stage of this work.

\end{document}